\documentclass{aa}

\usepackage{euclid}
\usepackage{graphicx}
\usepackage{natbib}
\usepackage{scalerel}

\usepackage{subfigure}
\usepackage{placeins}
\usepackage{sidecap}
\usepackage{siunitx}
\usepackage{calrsfs}
\usepackage{cprotect}
\usepackage[dvipsnames]{xcolor}

\newcommand{\X}[1]{\boldsymbol{X}_{#1}}
\newcommand{\q}[1]{\boldsymbol{q}_{#1}}
\newcommand{\qh}[1]{\breve{q}_{#1}}

\newcommand{\vth}[1]{\breve{\vartheta}_{#1}}
\newcommand{\ba}{\bb{a}}
\newcommand{\Ga}[1]{\Gamma_{#1}}
\newcommand{\g}{\gamma}
\newcommand{\m}[1]{\mathrm{#1}}
\newcommand{\mmmm}{\langle M^4 \rangle(\theta)}
\newcommand{\mmmms}{\langle M^*M^3\rangle(\theta)}
\newcommand{\mmmsms}{\langle M^{*2}M^{2}\rangle(\theta)}
\newcommand{\bvt}{\boldsymbol{\vartheta}}
\newcommand{\ulessqh}[1]{\breve{\mathscr{q}}_{#1}}

\newcommand{\bulessq}[1]{\boldsymbol{\mathscr{q}}_{#1}}

\newcommand{\bb}[1]{\boldsymbol{#1}}
\newcommand{\eiphi}[1]{\mathrm{e}^{#1 \ii \phi}}

\usepackage{txfonts}
\usepackage[pdfencoding=auto,psdextra]{hyperref}
\hypersetup{
    colorlinks=true,
    linkcolor=blue,
    filecolor=magenta,      
    urlcolor=blue,
    citecolor=blue
}
\makeatletter
\renewcommand*\aa@pageof{, page \thepage{} of \pageref*{LastPage}}
\makeatother

\usepackage[utf8]{inputenc}

\usepackage[switch, modulo]{lineno}
\usepackage{amsmath}	% Advanced maths commands
\usepackage{dsfont}
\usepackage{physics}
\usepackage{newtxtext,newtxmath}
\usepackage[export]{adjustbox}
\usepackage{tikz}
\usetikzlibrary{angles, quotes}
\usetikzlibrary{shapes.arrows,calc}
\usepackage{multirow}

\usepackage[margin=10pt,font=small,labelfont=bf]{caption} 

\newcommand{\Fig}[1]{Fig. \ref{#1}} %AandA
\newcommand{\secref}[1]{Sect. \ref{#1}}
\newcommand{\appref}[1]{Appendix \ref{#1}}
\newcommand{\Eq}[1]{Eq. (\ref{#1})}

\renewcommand{\ee}{\mathrm{e}}

\newcommand{\bvartheta}{\boldsymbol{\vartheta}}
\newcommand{\ii}{\mathrm{i}}

\newcommand{\Omegam}{\Omega_{\mathrm{m}}}

\newcommand{\ttheta}[1]{\pmb{\theta}}

\newcommand{\Norm}[2]{\mathcal{N}}

\newcommand{\map}{M_{\mathrm{ap}}}
\newcommand{\mperp}{M_{\times}}

\newcommand{\mapfour}{\map^4}

\newcommand{\mathitq}{\mathrm{\textbf{\textit{q}}}}

\newcommand{\npcf}{$n$PCF}

\tikzset{Pfeil/.style=%
{to path={let \p1 = ($(\tikztotarget)-(\tikztostart)$),
            \n1 = {int(mod(scalar(atan2(\y1,\x1))+360, 360))}, % calculate angle in range [0,360)
            \n2 = {veclen(\x1,\y1)}
        in \pgfextra{\typeout{\n1,\n2,\x1,\y1}}
         (\tikztotarget)
        node[draw,single arrow,
             minimum height=\n2-\pgflinewidth,
             inner sep=1ex,
             single arrow head extend=1ex,
             rotate=\n1, % not shape border rotate, because that for some reason didn't work
             anchor=tip, % anchor=tip added, pos=0.5 removed
             ]{}
         }}}

\begin{document}

\defcitealias{jarvis2004}{J04}
\defcitealias{porth2023}{P24}
\defcitealias{porth2025}{P25}

%%%%%%%%%%%%%%%%%%%%%%%%%%%%%%%%%%%%%%%%
%\usepackage[options]{hyperref}
% To add links in your PDF file, use the package "hyperref"
% with options according to your LaTeX or PDFLaTeX drivers.
%

   \title{Towards an application of fourth-order shear statistics I}

   \subtitle{The information content of $\langle M_\m{ap}^4 \rangle $}

   \author{Elena Silvestre-Rosello
          \inst{1,2}
          \and
          Lucas Porth\inst{2}
          \and
          Peter Schneider\inst{2}
          \and
          Laila Linke\inst{1}
          \and 
          Jonas Krueger\inst{1}
          \and 
          Sebastian Grandis\inst{1}
          \and 
          Jonathan Oel\inst{2}
          }

   \institute{Universit\"at Innsbruck, Institut für Astro- und Teilchenphysik, Technikerstr. 25/8, 6020 Innsbruck, Austria
\and University of Bonn, Argelander-Institut f\"ur Astronomie, Auf dem H\"ugel 71, 53121 Bonn, Germany\\
              \email{elena.silvestre-rosello@uibk.ac.at}
             }

   \date{September 8, 2025}

% \abstract{}{}{}{}{} 
% 5 {} token are mandatory
 
  \abstract
  % context heading (optional)
  % {} leave it empty if necessary  
   {Higher-order shear statistics contain part of the non-Gaussian information of the projected matter field and therefore can provide additional constraints on the cosmological parameters when combined with second-order statistics.}
  % aims heading (mandatory)
   {We aim to provide the theoretical framework for studying shear four-point correlation functions (4PCF) using fourth-order aperture statistics and develop a numerical integration pipeline to compute them. Finally, we forecast the information content of fourth-order aperture statistics.}
  % methods heading (mandatory)
   {We begin by giving the relation of the $n$-th order aperture statistics, $\langle M_\m{ap}^n \rangle$, to the shear $n$PCF and to the convergence polyspectra. We then focus on the fourth-order case, where we derive the functional form of their filters and test the behavior of these filters by numerically integrating over the 4PCF of a Gaussian random shear field (GRF). Finally, we perform a Fisher forecast on the constraining power of $\langle M_\m{ap}^4\rangle_\m{c}$, where we develop a novel method to estimate derivatives from a simulation suite with arbitrarily distributed cosmological sets.}
  % results heading (mandatory)
   {By analyzing and mitigating numerical effects within the integration pipeline, we achieve a two-percent-level precision on the fourth-order aperture statistics for a GRF, which remains well below the noise budget of Stage IV surveys. We report a minimal improvement in the constraining power of the aperture statistics when including fourth-order statistics to a $\langle M_\m{ap}^2\rangle + \langle M_\m{ap}^3\rangle$ joint analysis for a DES-Y3-like setup, using non-tomographic equal-scale aperture statistics.}
  % conclusions heading (optional), leave it empty if necessary 
   {}


   \maketitle
%
%-------------------------------------------------------------------
\section{Introduction}
For the past few decades, cosmic shear has provided insights into the contents and structure of the large-scale Structure (LSS) of the Universe. Second-order shear statistics, which capture the Gaussian information in the field, have been key in constraining the cosmological parameters. The results for Stage III surveys can be found in \citet{amon2021} and \citet{secco2021} for DES-Y3, \citet{wright2025} and \citet{stoelzner2025} for KiDS Legacy, or \citet{dalal2023} and \citet{Li2023} for HSC. As the observational precision increases, the significance of three-point statistics and other higher-order statistics also grows, successfully targeting the non-Gaussianities in the matter field, as forecasted by \citet{higher_order_euclid} on a \textit{Euclid} setup \citep{euclidoverview}. In Stage III surveys, both two- and three-point shear statistics have been explored, as their joint analysis yields tighter constraints on the cosmological parameters than either can provide individually, as shown in \citet{Burger2024} for a joint analysis using the Kilo Degree Survey \citep{kuijken2015}. 

The exceptional precision of stage IV surveys, like \textit{Euclid}, will give high signal-to-noise ratio (S/N) to statistics beyond third order \citep{higher_order_euclid}. This work is part of a two-part study, in which we explore fourth-order shear statistics and work toward their application. 

Fourth-order statistics may break degeneracies between second- and third-order statistics, giving therefore tighter constraints on the cosmological parameters. Alternatively, they may be degenerate with the previous orders, in which case they present a new test for systematics and study of nuisance parameters, such as intrinsic alignment or baryonic models. Both scenarios can be relevant for Stage IV surveys, where second- and higher-order statistics will become standard tools for cosmology.

We begin with a theoretical analysis on $n$-th order aperture statistics -- extending prior work on second-order \citep{schneider2002b} and third-order \citep{schneider2005, heydenreich2023} formulations -- and their connection to $n$-point polyspectra and $n$-point correlation functions ($n$PCFs). We then specialize to the fourth-order case, deriving the relation between fourth-order aperture statistics and the four-point correlation functions (4PCFs) of the shear field. Aperture statistics decompose the shear field into $E$- and $B$-modes, thus enabling robust detection of systematic errors. When computed via $n$PCFs, the aperture statistics are less sensitive to finite-field effects and masking biases \citep{jarvis2004,schneider2005,porth2020}. However, the numerical transformations between $n$PCFs and aperture statistics can introduce small biases, which we study and mitigate.

The companion paper, 
(\citealp[][hereinafter \citetalias{porth2025})]{porth2025}, presents an efficient estimator for the 4PCF based on a multipole decomposition, which computationally allows the measurement of fourth-order shear statistics. Building on the theoretical framework and computational pipelines established here, \citetalias{porth2025} report a significant detection of fourth-order aperture statistics using data from the first data release of the Dark Energy Survey (DES-Y3; \citealp{desy3}).

Finally, we assess the incremental constraining power of including fourth-order statistics alongside second- and third-order aperture statistics. In particular, we evaluate the information content of the fourth-order statistics $\langle M_\mathrm{ap}^4 \rangle$ with a Fisher forecast on a DES-Y3-like setup.

The structure of this paper is as follows: In \secref{sec:basic_concepts} we introduce the shear $n$PCF and the corresponding $n$-th order aperture statistics. Section \ref{sec:4thorder_measures} develops the fourth-order case analytically. In \secref{sec:grf} we numerically compute the aperture statistics using a model for a Gaussian random field and assess the accuracy of the computation. Section \ref{sec:forecast} presents a Fisher forecast on the constraining power of fourth-order aperture statistics in a DES-Y3-like setup. We summarize and discuss the main results in \secref{sec:discussion}.

%--------------------------------------------------------------------
\section{Higher-order measures of cosmic shear}
\label{sec:basic_concepts}
%--------------------------------------------------------------------
The fundamentals of weak lensing are described in \citet{bartelmannReview}, with a more recent cosmic shear description in \citet{kilbingerReview} and \citet{mandelbaumReview}.

\subsection{Basic concepts}
In cosmology, the matter density field is usually characterized by the density contrast
\begin{equation}
    \delta(\bb{X};\chi) = \frac{\rho_\m{m}(\bb{X};\chi)-\overline{\rho}_\m{m}(\chi)}{\overline{\rho}_\m{m}(\chi)},
\end{equation}
with $\rho_\m{m}(\bb{X};\chi)$ the three-dimensional matter density at an angular position, $\bb{X}$, and comoving distance, $\chi$; and $\overline{\rho}_\m{m}(\chi)$ its average value at a given $\chi$.

The dimensionless surface mass density or convergence, $\kappa(\bb{X};\chi)$, is defined as the projection of the three-dimensional matter field along the line of sight from the observer until a given $\chi$. For a flat universe it is given as
\begin{eqnarray}
\begin{aligned}
    \kappa&(\bb{X};\chi) =\frac{3H_0^2 \Omegam}{2c^2}\\
    &\times\int_0^{\chi} \diff \chi' \frac{(\chi-\chi')\chi'}{\chi} \, \delta\left(\chi'\bb{X},\chi\right)\ \left[1+z(\chi')\right],
    \label{eq:k_projected}
\end{aligned}
\end{eqnarray}
where $H_0$ is the Hubble constant, $\Omegam$ the matter density parameter, $c$ the speed of light, and $z$ the redshift, related to the comoving distance in the line of sight by $\diff z\,{c}/{H(z)}= \diff \chi$, with $H(z)$ the expansion rate of the Universe, from the first Friedmann equation \citep{friedmann1922}. In weak lensing, $\kappa(\bb{X};z)$ is a measure of the widening of light bundles at angular position $\bb{X}$ from sources at redshift $z$. If the sources follow a normalized redshift distribution, $p_z(z)$, which extends to a horizon redshift, $z_\m{h}$, then
\begin{equation}
    \kappa(\bb{X})= \int_0^{z_\m{h}} \diff z \ p_z(z) \, \kappa(\bb{X};z).
    \label{eq:kappa_pz}
\end{equation}

The shear $\g_\m{cart}(\bb{X};z)$ accounts for the distortion of such light bundles. It can be expressed as complex number with respect to a reference direction or projection, described by the angle $\zeta_i$ with respect to the Cartesian axis; for a given redshift
\begin{equation}
    \gamma(\bb{X}; \zeta_i) = - \g_{\mathrm{cart}}(\bb{X})\m{e}^{-2\m{i}\zeta_i} = \g_{\m{t}}(\bb{X};\zeta_i) + \m{i}\g_{\times}(\bb{X};\zeta_i),
\end{equation}
where we further decomposed the shear into its tangential and cross components, the value of which is projection dependent. Convergence and shear are related in Fourier space through
\begin{equation}
    \hat{\g}(\bb{\ell})= \hat{\kappa}(\bb{\ell})\m{e}^{2\m{i} \beta}\ \mathrm{and} \ 
    \hat{\g}^*(\bb{\ell})= \hat{\kappa}^*(\bb{\ell})\m{e}^{-2\m{i} \beta} = \hat{\kappa}(-\bb{\ell})\m{e}^{-2\m{i} \beta},
    \label{eq:kappagamma}
\end{equation}
with $\beta$ the polar angle of $\bb{\ell}$, as in \citet{kaiser1993}. 
%In the last step we used that $\kappa$ is a real field.

\subsection{Convergence $n$-point polyspectra and shear $n$-point correlation functions}
The $n$-point correlation function of the convergence ($\kappa n$PCF) describes the statistical correlation of the convergence field as a function of the angular distribution of $n$ points, and is defined as $\left\langle \prod_{i=0}^{n-1} \kappa(\X{i}) \right\rangle$. It is related to the $n$-point polyspectrum $\mathcal{P}_{\kappa,n}$ through the Fourier transform of the convergence, $\hat{\kappa}(\bb{\ell})$,
\begin{equation}
    \left\langle \prod_{i=1}^{n} \hat{\kappa}(\bb{\ell_i}) \right\rangle = \left( 2\pi \right)^2 \mathcal{P}_{\kappa,n} (\bb{\ell}_1, \cdots, \bb{\ell}_n)\ \delta_\m{D} \left(\sum_{i=1}^n \bb{\ell}_i\right).
    \label{eq:polyspectrum}
\end{equation}
Following from Eqs. (\ref{eq:k_projected}) and (\ref{eq:kappa_pz}), $\mathcal{P}_{\kappa,n}$ is a projection of the matter $n$-point polyspectrum $\mathcal{P}_{\delta,n}$ with the weighting function
\begin{equation}
    g(z) = \chi(z)\int_z^{z_\m{h}} \diff z_\m{s} \ p_z(z_\m{s})\frac{\chi(z_\m{s})-\chi(z)}{\chi(z_\m{s})},
\end{equation}
which accounts for the lensing efficiency of lenses at redshift $z$.\footnote{Some redshift information can be recovered by splitting the sources in tomographic bins with distinct $p_z(z_\m{s})$.} The projected polyspectrum then reads, under the Limber and flat-sky approximations \citep{limber1953},
\begin{eqnarray}
\begin{aligned}
    \mathcal{P}_{\kappa,n}&(\bb{\ell}_1, \cdots, \bb{\ell}_n) =
    \left(\frac{3H_0^2\Omega_\m{m}}{2 c^2}\right)^n\\
    &\times \int_0^{z_\m{h}} \diff z \frac{g^n(z)\left(1+z\right)^n}{\left[\chi(z)\right]^{2(n-1)}}
    \mathcal{P}_{\delta,n}\left(\frac{\bb{\ell}_1}{\chi(z)}, \cdots, \frac{\bb{\ell}_n}{\chi(z)},z\right).
    \label{eq:pk_from_pd}
\end{aligned}
\end{eqnarray}
The lowest order polyspectra are commonly referred to as the power spectrum ($n=2$), the bispectrum ($n=3$), and the trispectrum ($n=4$). 

\vspace{0.3cm}
An $n$-point correlation function of shear ($\gamma n$PCF or in the following simply $n$PCF) is defined as $\left\langle \prod_{i=0}^{n-1} \gamma^{(*)}(\X{i};\zeta_i) \right\rangle_p$, where $p$ is the number of points with conjugated shear, without fixing any specific order. These are related to the polyspectra using Eqs.\ (\ref{eq:kappagamma}) and (\ref{eq:polyspectrum}).

In order to compute an $n$PCF, the shear at every point $\bb{X}_i$ must be projected onto the direction determined by $\zeta_i$, defining an $n$PCF projection $\pi \equiv \{\zeta_1,\cdots,\zeta_n\}$. For the two-point case the shear is projected onto the direction between the given points, so the 2PCF depends only on the separation $\theta$ between them. The 2PCF can be decomposed in the quantities $\xi_+(\theta)$, $\xi_-(\theta)$, and $\xi_{\times}(\theta)$, where the first two are related to the convergence power spectrum as described in \citet{schneider2002b} and the last one is expected to vanish due to parity invariance \citep{schneider2003}.

Generalizing the prescription outlined in \citet{schneider2002}, the $n$PCF of a polar field, as the shear field, consists of $2^{n-1}$ independent complex functions called the natural components $\Gamma_{\mu}^{\pi,n},\ \mu=\{0,\cdots,2^{n-1}-1\}$. The order of the different natural components is described in Appendix G.1 in \citetalias{porth2025}, and specified for $n=4$ in \secref{sec:4PCF}. For $n>2$, the natural choice of projection $\pi$ is not obvious and can lead to coordinate singularities. The study of the 4PCF is our main focus and will be further discussed in \secref{sec:4PCF}.

\subsection{Aperture statistics}
The aperture mass statistics are a tool to decompose the shear field in its $E$- and $B$-modes by integrating the shear within a circular aperture of radius $\theta$, centred at a position $\bvt$. The $E$-mode, or pure lensing signal, is contained in the aperture mass $M_\m{ap}$ \citep{schneider1996}; while $M_\times$ quantifies the $B$-mode present in the field and can be used to estimate the mode-mixing due to higher-order and systematic effects \citep{schneider2002b}. Together they form the complex aperture measure
\begin{align}
    M(\bvt;\theta) &= M_\m{ap}(\bvt;\theta) + \m{i}M_\times(\bvt;\theta) \\
    &= \int \m{d} \vartheta' \, \vartheta' \, Q_\theta (\vartheta') \int \m{d} \zeta' \g(\bvt+\bvt';\zeta'),
    \label{eq:m}
\end{align}
with $\g$ the shear and $\zeta$ the radial direction of $\bvt'$. The aperture mass $M_\m{ap}$ can also be expressed as an integral over the convergence
\begin{eqnarray}
    M_\m{ap}(\bvt,\theta) = \int \m{d}^2\, \vt'\, U_\theta (|\bvt-\bvt'|)\, \kappa(\bvt'),
    \label{eq:map}
\end{eqnarray}
with the compensated filter $U_\theta (|\bvt|)$ related with the filter $Q_\theta(|\bvt|)$ by
\begin{equation}
    Q_\theta (\vt) = \frac{2}{\vt^2}\, \int_0^{\vt} \m{d} \vt'\, \vt' \, U_\theta (\vt')\, -\,  U_\theta (\vt).
\end{equation}

Here we use the Gaussian filter by \citet{crittenden2002},
\begin{align}
    \label{eq:filter_crit}
    Q_\theta (\vt) &= \frac{\vt^2}{4 \pi \theta^4}\exp{-\frac{\vt^2}{2\theta^2}}, \\
    U_\theta (\vt) &= \frac{1}{2\pi\theta^2}\left(1-\frac{\vt^2}{2\theta^2}\right)\exp{-\frac{\vt^2}{2\theta^2}}.
\end{align}

The moments of these complex aperture measures are the ensemble average of the product of apertures. The direct estimator for the moments, introduced by \citet{schneider1998} for second-order moments and extended i.e.\ by \citet{munshi2003}, computes the ensemble average as the average over all possible $N_\m{a}$ aperture centres,
\begin{equation}
    \left<  M_\m{ap}^n \right>_\m{dir} (\theta_1,\cdots,\theta_n) = \frac{\sum_i^{N_\m{a}} M_\m{ap}(\bvt_i,\theta_1)\cdots M_\m{ap}(\bvt_i,\theta_n)}{N_\m{a}},
    \label{eq:direct}
\end{equation}
where the individual aperture masses are computed from \Eq{eq:m} or \Eq{eq:map}. This estimator is biased when computed on masked fields, presents $E$- and $B$-mode mixing on finite fields \citep{heydenreich2023}, and requires the subtraction of multiple-counting terms \citep{porth2021}.

The moments of the complex apertures can also be expressed as integrals of the shear correlators over the $(2n)$-dimensional configuration space. For an $n$PCF with $p$ conjugated shears, where the shears are conjugated in the first $p$ positions,
\begin{eqnarray}
\begin{aligned}
    \left< (M^*)^p \right.& \left. M^{n-p} \right> (\theta_1,\cdots,\theta_n)= (-1)^n\int  \diff^2 X_1 \cdots \int  \diff^2 X_n\, \\
        &\times \left(\prod_{i=1}^{p} \,Q_{\theta_i} (|\bb{X_i}|) \frac{\breve{X}_i}{|\breve{X}_i|} \right)\left(\prod_{j=p+1}^{n} \,Q_{\theta_j} (|\bb{X_j}|) \frac{\breve{X}^*_j}{|\breve{X}_j|} \right)\\
        &\times \left<\prod_{i=1}^{p}\g_\m{cart}^{*}(\X{i})\prod_{j=p+1}^{n}\g_\m{cart}(\X{j})\right> \ ,
    \label{eq:m4_base}
\end{aligned}
\end{eqnarray}
where $\breve{X}_i$ is the complex representation of the vector $\bb{X}_i$. Equivalent expressions can be obtained for the $n$PCF with the $p$ conjuated shears in different positions. The shear correlator can be expressed in terms of the natural components of the $n$PCF, in which case we find $2^{n-1}$ distinct integrals, that are not affected by masks in the fields \citep{jarvis2004,schneider2005,porth2020}. In \appref{app:NthOrderMap4} we derive the expressions for the filters for equal-scale aperture radii (\ref{app:NthOrderMap4Same}), the generalization to multi-scale aperture radii (\ref{app:NthOrderMap4Diff}), and the separation into $E$- and $B$-modes (\ref{app:NthOrderSeparation}).
%The moments of the complex apertures can also be expressed as integrals over the $(2n)$-dimensional configuration space,
%\begin{eqnarray}
%\begin{aligned}
%    \left< \prod_{i=0}^{n-1} M^{(*)} \right>_p& (\theta_1,\cdots,\theta_n)= (-1)^n\int  \diff^2 X_1 \cdots \int  \diff^2 X_n\, \\
%        &\times \left(\prod_{i=1}^{n} \,Q_{\theta_i} (|\bb{X_i}|) \frac{\breve{X}^*_i}{|\breve{X}_i|} \right) \left<\prod_{i=0}^{n-1}\g_\m{cart}^{(*)}(\X{i})\right>_p \ ,
%    \label{eq:m4_base}
%\end{aligned}
%\end{eqnarray}
%where $p$ out of the $n$ different $M$ (and $\g$) are conjugated and $\breve{X}_i$ is the complex representation of the vector $\bb{X}$. The shear correlator can be expressed in terms of the $n$PCF, in which case we find $2^{n-1}$ distinct integrals, that are not affected by masks in the fields \citep{jarvis2004,schneider2005,porth2020}. In \appref{app:NthOrderMap4} we derive the expressions for the filters for equal-scale aperture radii (\ref{app:NthOrderMap4Same}), the generalization to multi-scale aperture radii (\ref{app:NthOrderMap4Diff}), and the separation into $E$- and $B$-modes (\ref{app:NthOrderSeparation}).

%Note that the shear correlator can be expressed in terms of the $n$PCF, in which case the $2^{n-1}$ distinct integrals and filters in \Eq{eq:m4_base} take the form in \appref{app:NthOrderMap4} for equal aperture radii and \appref{app:NthOrderMap4Diff} for different ones, with the $n$PCF as in Appendix G.1 in \citetalias{porth2025}. A separation in $E$- and $B$-modes to $n$-th order is possible as described in \appref{app:NthOrderSeparation}.

Alternatively, $\left<M_\m{ap}^n\right>$ can be expressed as an integral in Fourier space over the $n$-point polyspectrum,
\begin{eqnarray}
\begin{aligned}
    \left<  M_\m{ap}^n \right>& (\theta_1,\cdots,\theta_n) 
        =\int \frac{\diff^2 l_1}{(2\pi)^2} \cdots \int \frac{\diff^2 l_{n-1}}{(2\pi)^2}\ \left(\prod_{i=1}^{n-1}\left[\hat{u}(\theta_i|\bb{\ell}_i|)\right]\right)\ \\
    &\times\ 
        \hat{u}\left(\theta_n\left|\sum_{i=1}^{n-1}\bb{\ell}_i\right|\right)\ \mathcal{P}_{\kappa,n} \left(\bb{\ell}_1, \cdots, \bb{\ell}_{n-1}, -\sum_{i=1}^{n-1}\bb{\ell}_i\right),
    \label{eq:mapn_poly}
\end{aligned}
\end{eqnarray}
where we used \Eq{eq:polyspectrum} and $\hat{u}(\theta l)$ is the Fourier transform of $U_\theta(\vt)$.

%By comparison of the lhs in \Eq{eq:mapn_poly} with \Eq{eq:map} one can define the $n$-order \emph{direct estimator} of $\left<M_\m{ap}^n\right>$ as a higher-order generalization of \citet{schneider1998}. 

\vspace{0.3cm}

It is useful to further decompose the $n$PCF and the aperture statistics into their connected and disconnected part, where the connected part contains the additional information to the $n$-th order, and the disconnected part is a combination of the $m$-th orders, $m<n$ \citep{bernardeau2002}. 
%Doing so on the $n$PCF level is theoretically possible but in practice not recommended, as $n$PCFs are usually noise dominated TODO REF. 
Due to the isotropy of space, $\left<M_\m{ap}\right>$ approaches zero, such that the aperture statistics for $n\in\{2,3\}$ only contain connected terms, and the ones for $n=4$ only contain connected and Gaussian ($n=2$) terms, as discussed in \secref{sec:grf}; see \citet{porth2021} for a discussion on higher-orders.

%--------------------------------------------------------------------
\section{Fourth-order measures of cosmic shear}
\label{sec:4thorder_measures}

%--------------------------------------------------------------------
From now on, we focus on the 4PCF and the fourth-order aperture measures.
\subsection{Natural components of the 4PCF}
\label{sec:4PCF}

The shear 4PCF are 
\begin{equation}
    \left< \g^{(*)}(\X{0};\zeta_0)\, \g^{(*)}(\X{1};\zeta_1)\, \g^{(*)}(\X{2};\zeta_2)\, \g^{(*)}(\X{3};\zeta_3)\right>_p,
\end{equation}
where the shear on each point can be either conjugated or not, i.e.\,$\g^{*}(\X{i};\zeta_i) = \left[\g_{\m{t}}(\X{i};\zeta_i) - \m{i}\g_{\times}(\X{i};\zeta_i)\right]$.
We define the natural components of the shear 4PCF similarly to \citet{schneider2002}, where for notational simplicity we drop the index related to the fourth-order and the redshift dependency,
\begin{equation}
    \Ga{0}^{\m{cart}}(\bb{X}_0,\bb{X}_1,\bb{X}_2,\bb{X}_3)= \left< \g_\m{cart}(\bb{X}_0)\g_\m{cart}(\bb{X}_1)\g_\m{cart}(\bb{X}_2)\g_\m{cart}(\bb{X}_3)\right>,
    \label{eq:ga0}
\end{equation}
where `cart' represents the projection onto the Cartesian axis. The other natural components read, with the same components as $\Ga{0}^\m{cart}$,
\begin{eqnarray}
    \begin{aligned}
         &\Ga{1}^{\m{cart}}= \left< \g^*\g\g\g\right>, &&\Ga{2}^{\m{cart}}= \left< \g\g^*\g\g\right>, &&\Ga{3}^{\m{cart}}= \left< \g\g\g^*\g\right>, \\
         &\Ga{4}^{\m{cart}}= \left< \g\g\g\g^*\right>, &&\Ga{5}^{\m{cart}}= \left< \g^*\g^*\g\g\right>, &&\Ga{6}^{\m{cart}}= \left< \g^*\g\g^*\g\right>,\\
          &\Ga{7}^{\m{cart}}= \left< \g^*\g\g\g^*\right> \ .
    \end{aligned}
\end{eqnarray}

The shear $n$PCF in a projection $\pi$ can be projected onto some other projection $\pi'$ with the projection operator $P^{\pi, \pi'}$, for the different $\mu$ components,
\begin{equation}
     \Ga{\mu}^{\pi'}(\bb{X}_0, \cdots, \bb{X}_3) = P_{\mu}^{\pi, \pi'}(\bb{X}_0,\cdots,\bb{X}_3)\, \Ga{\mu}^{\pi}(\bb{X}_0,\cdots, \bb{X}_3).
     \label{eq:projection}
\end{equation}
Owing to the homogeneity and isotropy of space, for a projection tied to the relative position of the points, $\Ga{\mu}^{\pi}(\bb{X}_0,\bb{X}_1,\bb{X}_2,\bb{X}_3) = \Ga{\mu}^{\pi}\left(\vt_1, \vt_2, \vt_3, \psi_{12}, \psi_{23}\right)$, with $\vt_j$ the lengths from the vectors $\bvt_j = \X{j}-\X{0}$, $j=\{1,2,3\}$ (called radial coordinates hereinafter), and $\psi_{12}$ and $\psi_{23}$ the angles $\widehat{\bvt_1\bvt_2}$ and $\widehat{\bvt_2\bvt_3}$ (called angular coordinates), as defined in \Fig{fig:cross_scheme}. Together the relative positions of the four points form a four-sided polygon or quadrilateral.

\subsubsection{$\times$-projection of shear}
\begin{figure}[htbp]
    \centering
    \includegraphics[width=.999\linewidth,valign=t]{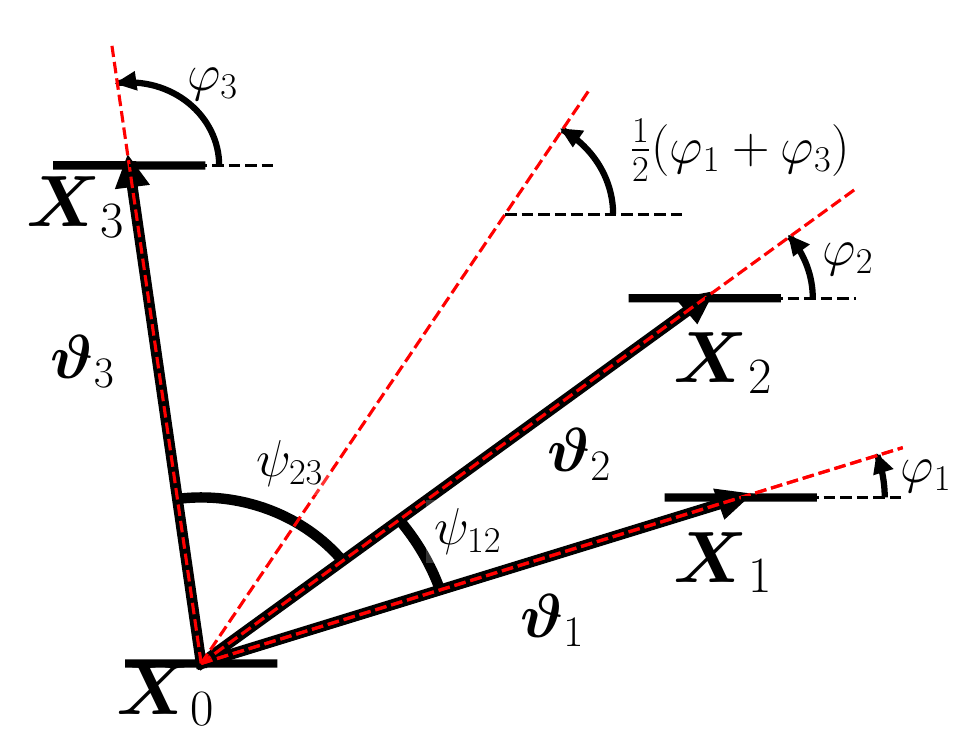}
    \caption{Parametrization of a quadrilateral using the $\times$-projection. Equivalent to Fig. (1) from \citetalias{porth2025} by defining $\phi_{12} = \psi_{12}$ and $\phi_{13} = \psi_{12} + \psi_{23}$.}
    \label{fig:cross_scheme}
\end{figure}

Here we consider the $\times$-projection for the shear, adapted from \citet{porth2023} and defined by the angles $\zeta^{\times}_j = \varphi_j$ for $j=\{1,2,3\}$ and $\zeta^{\times}_0 = (\varphi_1 + \varphi_3)/2$ from \Fig{fig:cross_scheme}. The projection operator for $\Ga{0}$ and $\Ga{1}$ from cartesian coordinates are, in terms of the complex quantities $\breve{\vt}_j = |\bvt_j|(\cos{\varphi_j}+\m{i}\sin{\varphi_j})$,
\begin{align}
    P_{0}^{\times,\m{cart}}\left(\bvt_1, \bvt_2, \bvt_3\right) &= \m{e}^{-2\m{i}\sum_{j=0}^3 \zeta^{\times}_j} = \m{e}^{-\m{i}\left(3\varphi_1+2\varphi_2 +3 \varphi_3\right)} \\&= \left(\frac{\breve{\vt}_1^*}{|\breve{\vt}_1|}\right)^3\left(\frac{\breve{\vt}_2^*}{|\breve{\vt}_2|}\right)^2\left(\frac{\breve{\vt}_3^*}{|\breve{\vt}_3|}\right)^3,\\
    P_{1}^{\times,\m{cart}} \left(\bvt_1, \bvt_2, \bvt_3\right)&= \m{e}^{-2\m{i}\left(-\zeta^{\times}_0 + \sum_{j=1}^3 \zeta^{\times}_j \right)}= \m{e}^{-\m{i}\left(\varphi_1+2\varphi_2 +\varphi_3\right)} \\ &=\left(\frac{\breve{\vt}_1^*}{|\breve{\vt}_1|}\right)\left(\frac{\breve{\vt}_2^*}{|\vth{2}|}\right)^2\left(\frac{\vth{3}^*}{|\vth{3}|}\right),
\end{align}
and analogous for all other natural components.

This projection is used for the multipole-based $n$PCF estimator described in \citetalias{porth2025}. Moreover, as shown in \secref{sec:grf}, the $\times$-projection also gives good results in transforming the 4PCF to the aperture measures. In practice, projecting means we replace the shear correlators in \Eq{eq:m4_base} by the product of $P_p^{\m{cart},\times}$ and the $n$PCF in the $\times$-projeciton.\footnote{A projection to the centre of mass of vertices (CMV) was also considered, where the shear on the $i$-th vertex is projected onto the vector $\bb{q_i}$ from the CMV to said vertex,
\begin{equation}
    \q{0} = -\frac{\bvt_1+\bvt_2+\bvt_3}{4}\ \m{and} \ \q{j} = \q{0} +\bvt_j
    %\q{j} = \frac{3\bvt_j-\bvt_{(j+1)\m{mod} 3}-\bvt_{(j+2)\m{mod} 3}}{4}
    , \  j = \{1,2,3\}.
    \label{eq:CMV_vectors}
\end{equation}
However, this projection is not used due to coordinate singularities for the filters when a quadrilateral vertex lays in the CMV.}

\subsection{Fourth-order aperture statistics}

We study the fourth-order aperture measures in the case of equal apertures, $\mmmm$, $\mmmms$, and $\mmmsms$, as \citet{heydenreich2023} and \citet{Burger2024} showed that the equal aperture case contain most of the information for the combined second- and third-order aperture statistics.

\subsubsection{Study on the filters for the 4PCF}

The 4PCF depends on the relative positions of the four points $\X{i}$, so the eight-dimensional integrals from \Eq{eq:m4_base} can be analytically reduced to five-dimensional ones by integrating over the position of the quadrilateral CMV and over the quadrilateral orientation $\varphi_1$; see Appendix \ref{sec:mmmm_derivation} for details.

The apertures measures related to $\Ga{0}$, $\Ga{1}$ and $\Ga{5}$ read (and similarly for $\Ga{2-4}$ and $\Ga{6-7}$)
\begin{align}
    \begin{split}
        &\mmmm = \prod_{j=1}^3 \left(\int  \frac{\diff \vt_j}{\theta}\, \frac{\vt_j}{\theta} \right) \int  \frac{\diff\psi_{12}}{2\pi} \int \frac{\diff \psi_{23}}{2\pi}\\
        &\times K_0\left(\frac{\q{0}}{\theta},\frac{\q{1}}{\theta},\frac{\q{2}}{\theta},\frac{\q{3}}{\theta}\right)\ P_{0}^{\m{cart},\times}\,\Ga{0}^\m{\times}(\vt_1, \vt_2, \vt_3 , \psi_{12},\psi_{23})\, ,\\
        &\mmmms =\prod_{j=1}^3 \left(\int  \frac{\diff \vt_j}{\theta}\, \frac{\vt_j}{\theta} \right) \int  \frac{\diff\psi_{12}}{2\pi} \int \frac{\diff \psi_{23}}{2\pi}\\
        &\times K_1\left(\frac{\q{0}}{\theta},\frac{\q{1}}{\theta},\frac{\q{2}}{\theta},\frac{\q{3}}{\theta}\right)\ P_{1}^{\m{cart},\times}\,\Ga{1}^\m{\times}(\vt_1, \vt_2, \vt_3 , \psi_{12},\psi_{23})\, ,\\
        &\mmmsms =\prod_{j=1}^3 \left(\int  \frac{\diff \vt_j}{\theta}\, \frac{\vt_j}{\theta} \right) \int  \frac{\diff\psi_{12}}{2\pi} \int \frac{\diff \psi_{23}}{2\pi}\\
        &\times K_5\left(\frac{\q{0}}{\theta},\frac{\q{1}}{\theta},\frac{\q{2}}{\theta},\frac{\q{3}}{\theta}\right)\ P_{5}^{\m{cart},\times}\,\Ga{5}^\m{\times}(\vt_1, \vt_2, \vt_3 , \psi_{12},\psi_{23})\ ,
        \label{eq:apertures_compute}
    \end{split}
\end{align}
where we defined the filter functions $K_m,\ m=\{0,\cdots,7\}$ in terms of the CMV vectors from \Eq{eq:CMV_vectors}.\footnote{\citetalias{porth2025} use the projection-dependent filter functions $$F_\mu^{(4,\times)}(\theta;\vartheta_1,\vartheta_2,\vartheta_3,\phi_{12},\phi_{13}) = K_\mu\left(\frac{\q{0}}{\theta},\cdots, \frac{\q{3}}{\theta}\right) P^{\m{cart},\times}_\mu (\bvt_1, \bvt_2, \bvt_3),$$ with $\q{i}$ from \Eq{eq:CMV_vectors}.} All dependencies of the filter on the aperture radius are through $\bvt/\theta$, so by defining $\bulessq{i} = \q{i}/\vt$,
\begin{align}
   \begin{split}
    K_0\left(\bulessq{0},\cdots,\bulessq{3}\right)=\frac{1}{64}\prod_{i=0}^3\left(\ulessqh{i}^{*2}\,\exp{-\frac{|\bulessq{i}|^2}{2}}\right), \label{eq:k0}
    \end{split}
\end{align} 
\begin{align}
   \begin{split}
    K_1\left(\bulessq{0},\cdots,\bulessq{3}\right)&=\frac{1}{64}\prod_{i=0}^3\left(\exp{-\frac{|\bulessq{i}|^2}{2}}\right)\times \left\{\,\ulessqh{0}^{2}\ulessqh{1}^{*2}\ulessqh{2}^{*2}\ulessqh{3}^{*2}+\vphantom{\frac{1}{2} \left(\ulessqh{1}^{*2}\left(\ulessqh{2}^{*2}+ 4 \ulessqh{2}^*\ulessqh{3}^*\right)+\m{2\ terms\ (1\rightarrow2\rightarrow3)})\right)}\right.\\
    &+2 \ulessqh{0}\left(\ulessqh{1}^{*}\ulessqh{2}^{*2}\ulessqh{3}^{*2}+\m{2\ cyclic}\right)+\\
    &\left.+\frac{1}{2} \left[\ulessqh{1}^{*2}\left(\ulessqh{2}^{*2}+ 4 \ulessqh{2}^*\ulessqh{3}^*\right)+\m{2\ cyclic}\right]\right\}\label{eq:k1}
    \end{split},
\end{align}
where the two terms correspond to a cyclic permutation of the indices 1 to 3, and
\begin{align}
   \begin{split}
    K_5\left(\bulessq{0},\cdots,\bulessq{3}\right)&=\frac{1}{64}
    \left(\prod_{i=0}^3\exp{-\frac{|\bulessq{i}|^2}{2}}\right)\times 
    \left[\,\ulessqh{0}^{2}\ulessqh{1}^{2}\ulessqh{2}^{*2}\ulessqh{3}^{*2}+
    \vphantom{3\left(\ulessqh{3}+\ulessqh{0}\right) \left(\ulessqh{1}^*+\ulessqh{2}^*\right) +\frac{3}{2}}\right.\\
    &+2\ulessqh{0}\ulessqh{1}\ulessqh{2}^*\ulessqh{3}^*\left(\ulessqh{0}+\ulessqh{1}\right)\left(\ulessqh{2}^*+\ulessqh{3}^*\right) +\\
    &+\frac{1}{2}\left(\ulessqh{0}^{2}+\ulessqh{1}^{2}+4\ulessqh{0}\ulessqh{1}\right) \left(\ulessqh{2}^{*2}+\ulessqh{3}^{*2}+4\ulessqh{2}^*\ulessqh{3}^*\right)
     +\\
    &+\left. 3\left(\ulessqh{0}+\ulessqh{1}\right)\left(\ulessqh{2}^*+\ulessqh{3}^*\right)  +\frac{3}{2}  \right].\label{eq:k5}
    \end{split}
\end{align}
The filter functions for the other natural components can be obtained from the ones shown above by renaming the vertex with a conjugated shear, i.e.
\begin{align}
   \begin{split}
    K_2(\ulessqh{0}, \ulessqh{1}, \ulessqh{2}, \ulessqh{3}) = K_1 (\ulessqh{1},\ulessqh{0}, \ulessqh{2}, \ulessqh{3}), \\
    K_3(\ulessqh{0}, \ulessqh{1}, \ulessqh{2}, \ulessqh{3}) = K_1 (\ulessqh{2}, \ulessqh{1}, \ulessqh{0}, \ulessqh{3}), \\
    K_4(\ulessqh{0}, \ulessqh{1}, \ulessqh{2}, \ulessqh{3}) = K_1 (\ulessqh{3}, \ulessqh{1}, \ulessqh{2}, \ulessqh{0}), \\
    K_6(\ulessqh{0}, \ulessqh{1}, \ulessqh{2}, \ulessqh{3}) = K_5 (\ulessqh{0}, \ulessqh{2}, \ulessqh{1},\ulessqh{3}), \\
    K_7(\ulessqh{0}, \ulessqh{1}, \ulessqh{2}, \ulessqh{3}) = K_5 (\ulessqh{0}, \ulessqh{3}, \ulessqh{2},\ulessqh{1}).
    \end{split}
    \label{eq:trafo_filter}
\end{align}

These filters are equivalent to those in \citet{jarvis2004}, considering the notation change in the natural components. A cyclic permutation of indices as described there is equivalent to the vertex exchange due to the symmetry of the filters.

Appendix \ref{sec:filters_plot} shows the dependency of the filters on the radial and angular coordinates. Due to their complicated profile, it is expected that the integration of the 4PCF over them requires some careful study, for which we refer to \secref{sec:grf}.

\subsubsection{Aperture statistics from 4PCF}
\label{sec:map4_from_m4}
An $E$/$B$-mode separation is possible following the prescription in \appref{app:NthOrderSeparation}, giving
\begin{align}
    \langle M_{\m{ap}}^4 \rangle =& \frac{1}{8} \Re \left[\left( 1,1,1,1,1,1,1,1\right)\cdot \mathcal{M}\right], \label{eq:map4p}\\
    \langle M_{\m{ap}}^3 M_\times \rangle=& \frac{1}{8} \Im \left[ \left( 1,\frac{1}{2},\frac{1}{2},\frac{1}{2},\frac{1}{2},0,0,0\right)\cdot \mathcal{M}\right],\\
    \langle M_{\m{ap}}^2 M_\times^2 \rangle =& \frac{1}{8} \Re\left[  \left( -1,0,0,0,0,\frac{1}{3},\frac{1}{3},\frac{1}{3}\right) \cdot \mathcal{M}\right],\\
    \langle M_{\m{ap}} M_\times^3 \rangle =& \frac{1}{8} \Im \left[ \left( -1,\frac{1}{2},\frac{1}{2},\frac{1}{2},\frac{1}{2},0,0,0\right)\cdot \mathcal{M}\right], \\
    \langle M_\times^4 \rangle =& \frac{1}{8}\Re  \left[\left( 1,-1,-1,-1,-1,1,1,1\right) \cdot\mathcal{M}\right],
\end{align}
with $\mathcal{M}= \left( \langle M^4 \rangle, \langle M^*M^3 \rangle, \langle MM^*M^2 \rangle,   \langle M^2M^*M\rangle,\langle M^3M^* \rangle, \right.$ $ \left.\langle M^{*2}M^{2} \rangle, \langle M^*MM^*M \rangle, \langle M^*M^2M^* \rangle \right)^{\m{T}}$, where all measures depend on $\theta$. The eight complex apertures measures are necessary for a decomposition of the fields in $E$- and $B$-modes in the presence of shape noise and tomographic measurements. 

For a pure $E$-mode field, $\langle M_{\m{ap}}^4 \rangle$ will be the only non-zero component, while  $\langle M_{\m{ap}}^2 M_\times^2 \rangle $  will be non-zero in the presence of $E$- and $B$-modes, and $\langle M_{\m{ap}}^3 M_\times \rangle$ and $\langle M_{\m{ap}} M_\times^3 \rangle$ will be non-zero for a parity violating field \citep{schneider2003}.

While the $E$/$B$-decomposition described above is exact theoretically, it cannot be realised for a measurement of the shear 4PCF on real data, which will be a noisy estimate of the bin-averaged 4PCF on a finite radial interval. We discuss the effects arising from the numerical integration in \secref{sec:grf}.

\subsubsection{Aperture statistics from trispectrum}
The fourth-order aperture mass can be obtained alternatively as an integral over the convergence trispectrum, specifying \Eq{eq:mapn_poly} to fourth order and equal-scale apertures,
\begin{eqnarray}
\begin{aligned}
    \left<M_\m{ap}^4\right>&(\theta)     
        =\int \frac{\diff^2 l_1}{(2\pi)^2} \cdots \int \frac{\diff^2 l_{3}}{(2\pi)^2}\ \prod_{i=1}^{3}\left[\hat{u}(\theta|\bb{\ell}_i|)\right]\\\
    &\times\hat{u}\left(\theta\left|\sum_{i=1}^{3}\bb{\ell}_i\right|\right)\ \mathcal{P}_{\kappa,4} \left(\bb{\ell}_1, \cdots, \bb{\ell}_{3}, -\sum_{i=1}^{3}\bb{\ell}_i\right).
\end{aligned}
\end{eqnarray}

\begin{figure}[htbp]
    \centering
    \includegraphics[width=.999\linewidth,valign=t]{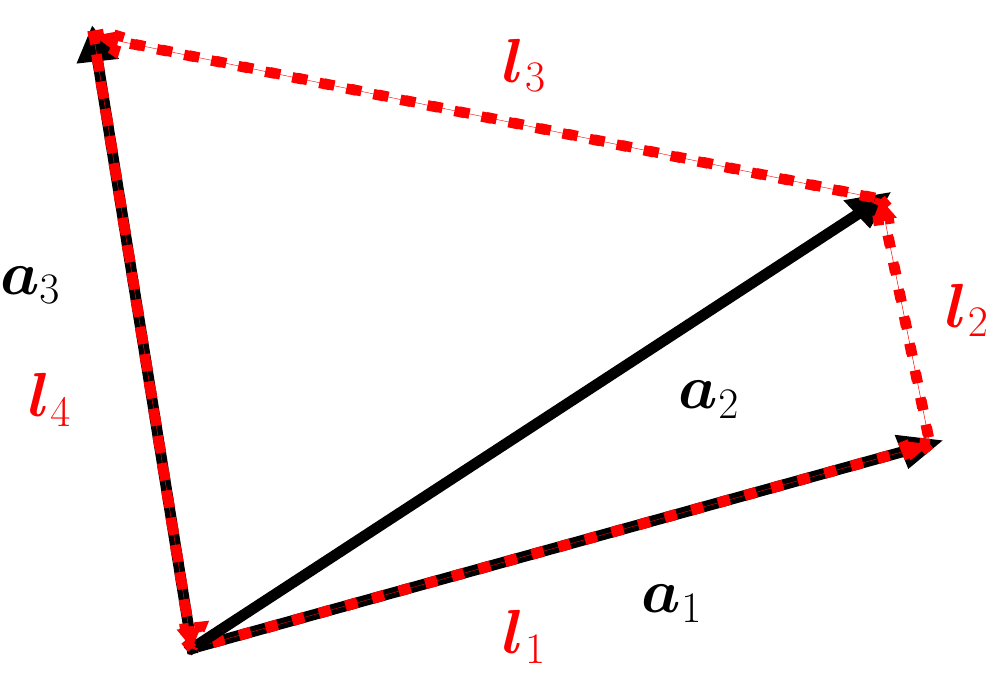}
    \caption{Sketch of the two sets of coordinates for the projected trispectrum.}
    \label{fig:sketch_tri}
\end{figure}

In Fourier space, the vectors $\bb{\ell}_\alpha,\ \alpha=\{1,\cdots,4\}$, form a closed quadrilateral, ensured by the Delta function from \Eq{eq:polyspectrum}. On the computational side, we find more convenient to change the integration coordinates to the vectors $\{\ba_1,\ba_2,\ba_3\} = \{\bb{\ell}_1,\bb{\ell}_1+\bb{\ell}_2, \bb{\ell}_1+\bb{\ell}_2+\bb{\ell}_3\}$ as defined in \Fig{fig:sketch_tri}, where the closed quadrilateral condition is directly satisfied,
\begin{eqnarray}
\begin{aligned}
    \left<M_\m{ap}^4\right>(\theta) =   &  
        \int \frac{\diff^2 a_1}{(2\pi)^2} \cdots \int \frac{\diff^2 a_{3}}{(2\pi)^2}\\
        & \times \mathcal{P}_{\kappa,4} \left(\ba_1, \ba_2-\ba_1,\ba_3-\ba_2, -\ba_3\right)\\
        & \times\
        \hat{u}(\theta|\ba_1|)\ \hat{u}(\theta|\ba_2-\ba_1|)\ \hat{u}(\theta|\ba_3-\ba_2|)\ \hat{u}(\theta|-\ba_3|) \\
        =&8\int_\mathbb{R} \frac{\diff a_1}{2\pi} \cdots \int_\mathbb{R} \frac{\diff a_{3}}{2\pi}\int_0^{\pi}\frac{\diff \alpha_{12}}{2\pi}\int_0^{\alpha_{12}}\frac{\diff \alpha_{23}}{2\pi}\\
        &\times \mathcal{P}_{\kappa,4} \left(\ba_1, \ba_2-\ba_1,\ba_3-\ba_2, -\ba_3\right)\\
        & \times\
        \hat{u}(\theta|\ba_1|)\ \hat{u}(\theta|\ba_2-\ba_1|)\ \hat{u}(\theta|\ba_3-\ba_2|)\ \hat{u}(\theta|-\ba_3|)\\
        & \times \ 
        \left(1-\frac{\delta_\m{D}(a_2-a_1)}{2}\right)
        \left(1-\frac{\delta_\m{D}(a_3-a_2)}{2}\right)\\
        &\times
        \left(1-\frac{\delta_\m{D}(a_1-a_3)\left[1-\delta_\m{D}(a_2-a_1)\right]}{2}\right)
\end{aligned}
\end{eqnarray}
where in the second step we reduce the integration hypervolume by considering that the quadrilateral is symmetric under exchange $\alpha_{12}\rightarrow-\alpha_{12}$ and $\alpha_{12}\leftrightarrow\alpha_{23}$. The $\delta_\m{D}$ factors avoid double counting in case two of the $a_i$ have the same value, the last one accounting for $(a_1 = a_2\ \&\ a_2 = a_3) \implies (a_1 = a_3)$.

However, we do not recommend this approach on observational data because the conversion from observable (real) space to Fourier space involves an integral over the whole real space, which is not practical due to finite survey sizes. Moreover, the integral over the $n$PCF is unaffected by masks in the field \citep{schneider2005,jarvis2004}.
%-------------------------------------------------------------------
\section{Case of Gaussian random fields}
\label{sec:grf}

%--------------------------------------------------------------------
We test the expressions for the aperture filters for the 4PCF, Eqs.\ (\ref{eq:k0})--(\ref{eq:k5}), by assuming a Gaussian random shear field (GRF), which is chosen for three main reasons. Firstly, GRF are fully described by their second-order moments, i.e. their 2PCFs or their power spectrum. Therefore, for such fields, any $(2n)$-PCF can be written as combination of 2PCFs, and any $(2n+1)$-PCF vanishes. In Appendix \ref{sec:4PCFas2PCF} we explicitly show the form of the shear 4PCFs in terms of the shear 2PCFs for a GRF. Secondly, in this case the fourth-order aperture statistics can also be decomposed into second-order aperture statistics, as we will discuss in \secref{sec:map4map2}. Finally, the shear field approaches, on large scales, a GRF, so on those scales the full 4PCF (connected and disconnected terms) is expected to behave similarly to the obtained from a realistic 2PCF, as seen in Fig. 3 from \citetalias{porth2025}.

\subsection{Fourth-order filters in terms of second-order filters}
\label{sec:map4map2}

In Appendix \ref{sec:m4asm2} we show the expressions of $\langle M^{(*)} M^{(*)} M^{(*)} M^{(*)} \rangle (\theta)$ as a function of $\langle M M \rangle (\theta)$ and $\langle M M^{*} \rangle (\theta)$ for the non-tomographic case and equal aperture scales. This allows us to relate the fourth- and second-order aperture statistics by using \Eq{eq:m},
\begin{align}
     \langle M_{\m{ap}}^4\rangle &=  3\left[ \langle M_{\m{ap}}^2\rangle\right]^2, \label{eq:map4asmap2}\\
     \langle M_{\m{ap}}^3M_\times\rangle &=3 \langle M_{\m{ap}}M_{\times}\rangle\langle M_{\m{ap}}^2\rangle,\\
     \langle M_{\m{ap}}^2M_\times^2\rangle &= \langle M_{\m{ap}}^2\rangle \langle M_{\times}^2\rangle + 2\left[\langle M_{\m{ap}}M_{\times}\rangle\right]^2,\\
    \langle M_{\m{ap}}M_\times^3\rangle &= 3 \langle M_{\m{ap}}M_{\times}\rangle\langle M_{\times}^2\rangle,\\
    \langle M_{\times}^4\rangle &= 3\left[ \langle M_{\times}^2\rangle\right]^2\label{eq:mx4asmx2},
\end{align}
where all quantities depend on $\theta$. For a general field, these describe the disconnected part of the fourth-order statistics, with $\langle M_{\m{ap}}^4\rangle$ the only non-zero component for a pure $E$-mode field.

\subsection{Integration routine}
\label{sec:integration}

\begin{figure*}[tbp]  
    \centering
    \includegraphics[width=1\textwidth]{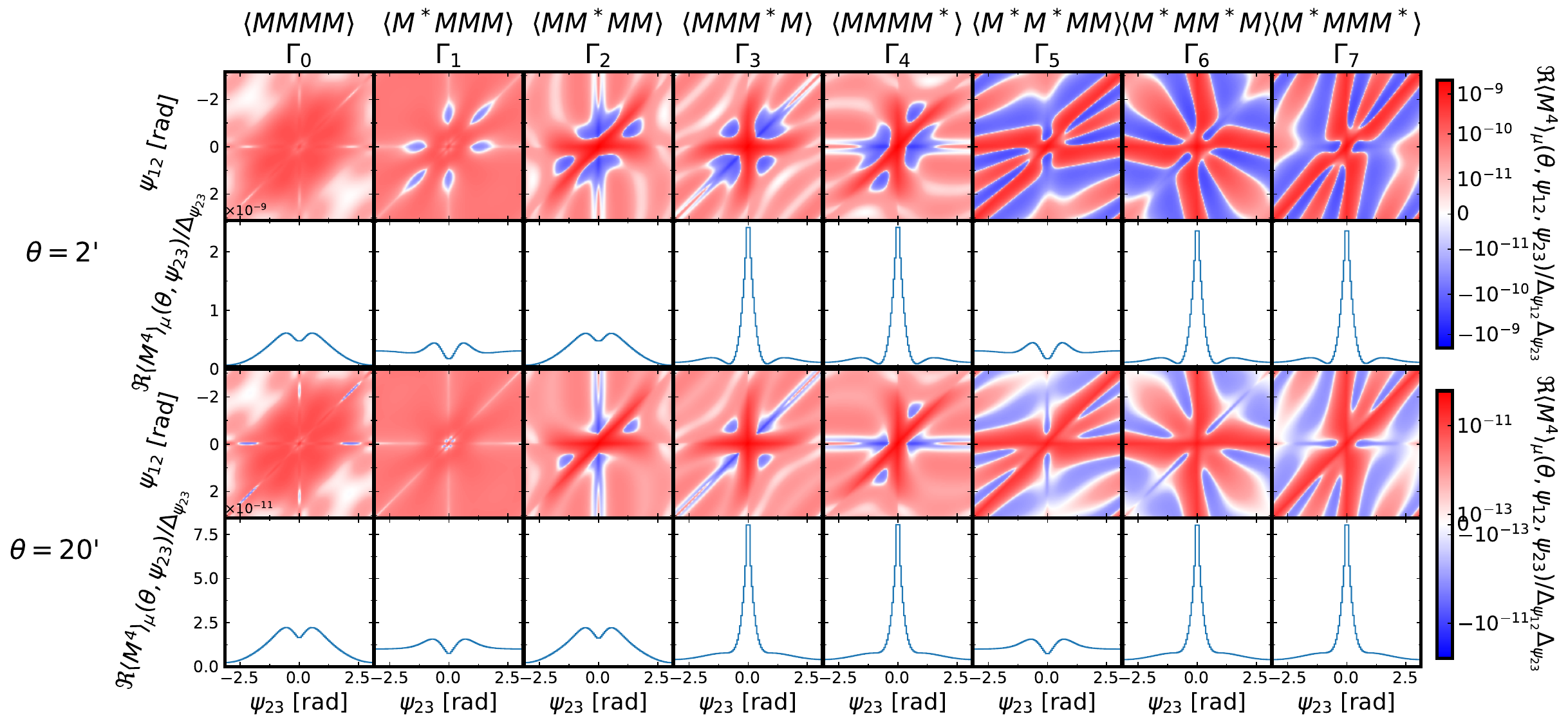}
    \cprotect\caption{Integrand for the eight complex
    aperture measures once the radial integration is performed (first and third rows) and once all but one angular integral are performed (second and fourth rows), 
    computed on 126 radial bins and 113 angular bins.
    At small aperture radii, $\theta = 2\,'$ (top two rows),
    we find a higher signal, while at large aperture radii, $\theta = 20\,'$ (bottom
    two rows), the signal is more strongly peaked around the zero-angles. The peaks from the one-dimensional profiles correspond to
    close-by mixed-pairs configurations, which play a stronger role for large $\theta$, therefore the smaller region of relative high values. Similar peaks as in the integral over
    $\Ga{2}$, $\Ga{3}$, $\Ga{6}$, and $\Ga{7}$ can be found for the integral over $\Ga{4}$ and $\Ga{5}$ by projecting onto another angle ($\psi_{23}$ or $\psi_{12}+\psi_{23}$).
    The integrals over $\Ga{0}$ and $\Ga{1}$ are smoother, so a faster convergence with binning accuracy is expected. The colour bars are in \verb|Symlog| scale \citep{matplotlib}, i.e.\ in logarithmic scale for large absolute values and linear scale for values around 0,
    with the linear threshold 170 times lower than the maximum for each aperture.}
    \label{fig:radial_integration}
\end{figure*}
The integrals in \Eq{eq:apertures_compute} are numerically computed using a model for a Gaussian shear field. For that, the 2PCFs are obtained with \verb|pyccl| \citep{pyccl},\cprotect\footnote{\url{https://github.com/LSSTDESC/CCL}} assuming a flat $\Lambda$CDM cosmology
with matter density $\Omega_\m{m}=0.30$, null baryonic and radiation density $\Omega_\m{b}=0.00=\Omega_\m{r}$, a reduced Hubble
constant $h = 0.7$,  $\sigma_8 = 0.8$, and spectral index $n_s = 0.96$. The shear field presents a Gaussian redshift distribution of mean $\overline{z}=1.5$
and standard deviation $\sigma_z = 0.15$. The 4PCFs are obtained from the linearly interpolated 2PCFs using the prescription in Appendix \ref{sec:4PCFas2PCF}. %The 4PCF obtained with this method are in the Cartesian projection and therefore depend on eight variables and cannot be measured. In order to avoid the conversion to $\times$ projection and the correspondent projector operator in \Eq{eq:apertures_compute}, we define $\Ga{i}^{\hat{\m{cart}}}(\vt_1, \vt_2, \vt_3, \psi_{12}, \psi_{23}) = \left.\Ga{i}^{\m{cart}}(\X{0}, \X{1}, \X{2}, \X{3})\right|_{\{\X{0}=(0,0),\X{1}=(\vt_1,0)\}}$.

We estimate \Eq{eq:apertures_compute} using a Riemann sum with $N_{\m{ang}}$ almost linearly spaced angular bins
for $\psi_{12}$ and $\psi_{23}$, and $N_\m{rad}$ logarithmically spaced radial bins
for $\vt_i$ (see \secref{sec:binning} for a discussion on the geometry). The integral over
the $l$-th natural component corresponds to the aperture measure $\left<M^{(*)4}\right>_\m{Rie}^{l}(\theta)$,
\begin{equation}
    \begin{aligned}
    \left<M^{(*)4}\right>_\m{Rie}^{l}&(\theta) = \sum_{i, j, k =1}^{N_\m{rad}}\frac{\vt_{i}\Delta_{\vt;i}}{\theta^2} \frac{\vt_{j} \Delta_{\vt;j}}{\theta^2}\frac{\vt_{k} \Delta_{\vt;k}}{\theta^2} 
     \\
     &\times\ \sum_{m, n = 1}^{N_{\m{ang}}}\frac{\Delta_{\psi;m}}{2\pi} \frac{\Delta_{\psi;n}}{2\pi}\ K_l  \left(\frac{\q{0}}{\theta},\frac{\q{1}}{\theta},\frac{\q{2}}{\theta},\frac{\q{3}}{\theta}\right)
     \\
     &\times\   P^{\m{cart},\times}_l\Ga{l}^{\times} \left(\vt_{i}, \vt_{j}, \vt_{k}, \psi_{12;m}, \psi_{23;n}\right) \ ,
    %&\times K_l  \left(\frac{\q{0;i,j,k,m,n}}{\theta},\frac{\q{1;i,j,k,m,n}}{\theta},\frac{\q{2;i,j,k,m,n}}{\theta},\frac{\q{3;i,j,k,m,n}}{\theta}\right)  ,
    \label{eq:thesum}
    \end{aligned}
\end{equation}
where $\vt_{i}$ is the arithmetic centre of the $i$-th radial bin of width $\Delta_{\vt;i}[\m{arcmin}]$,  $\psi_{12;m}$ is the arithmetic centre of the
$m$-th angular bin of width $\Delta_{\psi;m}[\m{rad}]$ and similarly for the other coordinates. The vectors $\q{0}$ to $\q{3}$ are computed from
\Eq{eq:CMV_vectors} using the centres of the binned radial and angular components.

The complexity of this method scales as $N_{\m{ang}}^2 N_{\m{rad}}^3$, which is computationally expensive. Nonetheless, as we show in the following, this is not a serious concern, since the bottleneck in the
application of fourth-order aperture statistics lays in the computation of the 4PCF, described in \citetalias{porth2025}.

In order to assess the precision of the estimated fourth-order aperture statistics, $\left<M^{(*)4}\right>_\m{Rie}^{l}$, we compare the estimate of $\left<M_\m{ap}^4\right>$ from \Eq{eq:map4p} with that of \Eq{eq:map4asmap2}, with $\left< M_\m{ap} ^2 \right>$ computed from the 2PCF using the transformation equations, Eq.\ (32) of \citet{schneider2002b} and the second-order filters in \citet{jarvis2004}.

\subsubsection{Bin geometry}
\label{sec:bin_geometry}
We find that not all quadrilateral configurations contribute equally to the given aperture measure. 
\Fig{fig:radial_integration} shows the integrand of \Eq{eq:thesum} after the radial integration for apertures with radii
 $\theta \in \{2\,', 20\,'\}$, in terms of the two angles from \Fig{fig:cross_scheme}. The integrand's one-dimensional angular profile is also shown in \Fig{fig:radial_integration} as the projection onto the $\psi_{12}$ axis. %\Fig{fig:radial_integration} shows as well the
%projection onto $\psi_{12}$ to study the integrand’s one-dimensional angular profile.

Most of the weight in the integrals is in quadrilateral configurations where at least two points are close by and are ``mixed pairs'', i.e.\ only one of the points has conjugated shear, so the phase factor in Eqs. (\ref{eq:Gamma0cart})--(\ref{eq:Gamma5cart}) is unity, an effect that is stronger for large aperture radii. The vertices with conjugated shear depend on the natural component over which we integrate. For $\Ga{1}$, for example, the conjugated shear
is on the vertex at $\X{0}$, so ``mixed pairs'' include such vertex and most power is around small $\vt_{i}$, with low angular dependency. For the integrals with $\Ga{i},\,i>1$, we find ``mixed pairs'' around vertices that are not in the origin, so most of the power comes from under- and asymmetrically-sampled regions around $\X{j},\ j =\{1,2,3\}$.

Due to the sharp peaks around the configurations with angles equal to zero, we choose an angular binning scheme that samples the angles $\psi_{12}$, $\psi_{23}$, and $(\psi_{12}+\psi_{23})$ symmetrically around $0\,\mathrm{rad}$. This will result in consistent underestimations for low binning accuracy, as seen in \secref{sec:binning}. Moreover, we find that the most consistent results are found when using hypercubical bins, where for each radial and angular coordinate $\vt\Delta_\psi\approx \Delta_\vt$, which in the limit of infinitesimally small bins $\Delta_\vt = \vt\Delta(\ln \vt)$ corresponds to $\Delta_\psi = \Delta(\ln \vt)$, i.e.\ logarithmically spaced radial bins.

%TODO ALTERNATIVE INTEGRATION SCHEMES?

\subsubsection{Integration range effects}
\label{sec:integration_range}
In a realistic case, the radial integration cannot extend over the entire domain due to a finite number density of sources and a finite survey extent. 

\begin{figure}[tbp]
    \centering
    \includegraphics[width=.999\linewidth,valign=t]{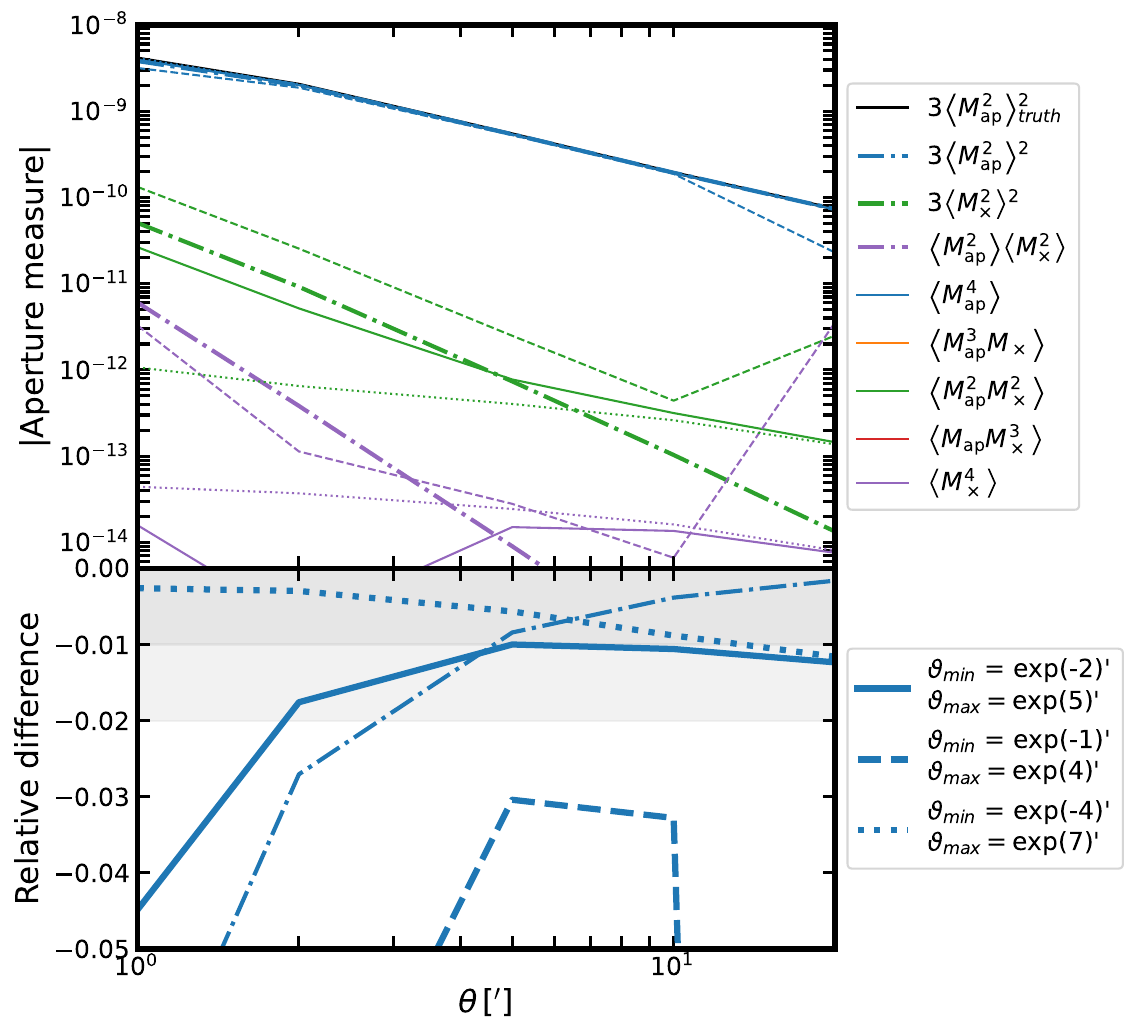}
    \caption{Effect of a finite radial integration range on the fourth-order aperture statistics. (Top) Aperture statistics for varying radial integration ranges, computed on a grid with a logarithmic bin width of $\Delta(\ln \vt) = 0.07$ and
    $\Delta_\psi = 0.07\,\m{rad}$m alongside with the considered ground truth (see text).
    Dash-dotted lines show the effect of $E$- and $B$-mode mixing in the second-order aperture statistics for a radial integration range $\vt\in[\m{e}^{-2}\,', \m{e}^5\,']$,
    propagated to the fourth-order with Eqs.\ (\ref{eq:map4asmap2})--(\ref{eq:mx4asmx2}). Parity violating modes are, in all cases considered here, below $10^{-20}$. (Bottom) Relative difference for $\left<M_\m{ap}^4\right>(\theta)$
    to the considered ground truth. The dash-dotted line shows the
    impact of a finite integration range on $3\left<M_\m{ap}^2\right>^2$, and the other lines on $\left<M_\m{ap}^4\right>$. Gray shaded regions show the $1\%$ and $2\%$ regions.}
    \label{fig:integration_range}
\end{figure}

%At $\vt < 0.2\,\m{arcmin}$ not many quadruplets are found, so a lower bound is set, below which we find low statistics.\footnote{In data from observations, the finite resolution of telescopes leads to source blending at very small scales, $\vt \lesssim 0.1\,\m{arcmin}$ for stage III surveys, which is not considered here.}
%Thanks to our choice of the filter function in the aperture statistics, it is not needed to integrate radially to infinity. An upper bound is chosen toreduce the required computational complexity.
On aperture radii $\theta \in [2\,', 20\,']$ an integration range $\vt\in[\vt_\m{min},\vt_\m{max}]=[\m{e}^{-2}\,', \m{e}^5\,']$ gives a result compatible with the expectation from second-order statistics on a few percent level, as shown in \Fig{fig:integration_range} for the preferred
set of binning parameters discussed in \secref{sec:binning}. Here and in Figs.\ \ref{fig:binning}--\ref{fig:integrated} we compute the relative difference with respect to a ground truth, estimated as $3\left<M_\m{ap}^2\right>^2$, computed with
integration range $\vt \in [\m{e}^{-5}\,', \m{e}^9\,']$ and a logarithmic bin width of $\Delta(\ln \vt)=0.04$.

The discrepancy for small integration ranges is dominated by an \textit{E}-mode/\textit{B}-mode mixing, arising from the impact of a finite integration range in the second-order aperture statistics described in \citet{kilbinger2006}. The brown lines in \Fig{fig:integration_range} show honw these effects propagate into the Gaussian fourth-order statistics through Eqs.\ (\ref{eq:map4asmap2})-(\ref{eq:mx4asmx2}), when the integration range is too small, here $\vt\in[\m{e}^{-2}\,', \m{e}^5\,']$. 

The behavior when $\left<M_\m{ap}^4\right>$ is estimated from the 4PCF with \Eq{eq:map4p} differs due to our integration scheme, since the $\vt$ boundaries restrict the minimum and maximum separation between pairs that include the vertex $\X{0}$, but the distances $d$ between other vertices depend on the bin distribution\footnote{We added an artificial boundary of $d\in [10^{-3}\,', 10^4\,']$ on the 2PCF}. On small scales the recovered $B$-mode is smaller than the one propagated from the 2PCF because, for some configurations, $d<\vt_\m{min}$. On large scales the obtained $B$-mode is larger due to the finite size of the logarithmic bins, that sets  $d>\vt_\m{max}$ for large $\vt_i$.

\subsubsection{Binning accuracy}
\label{sec:binning}

\begin{figure}[htbp]
    \centering
    \includegraphics[width=.999\linewidth,valign=t]{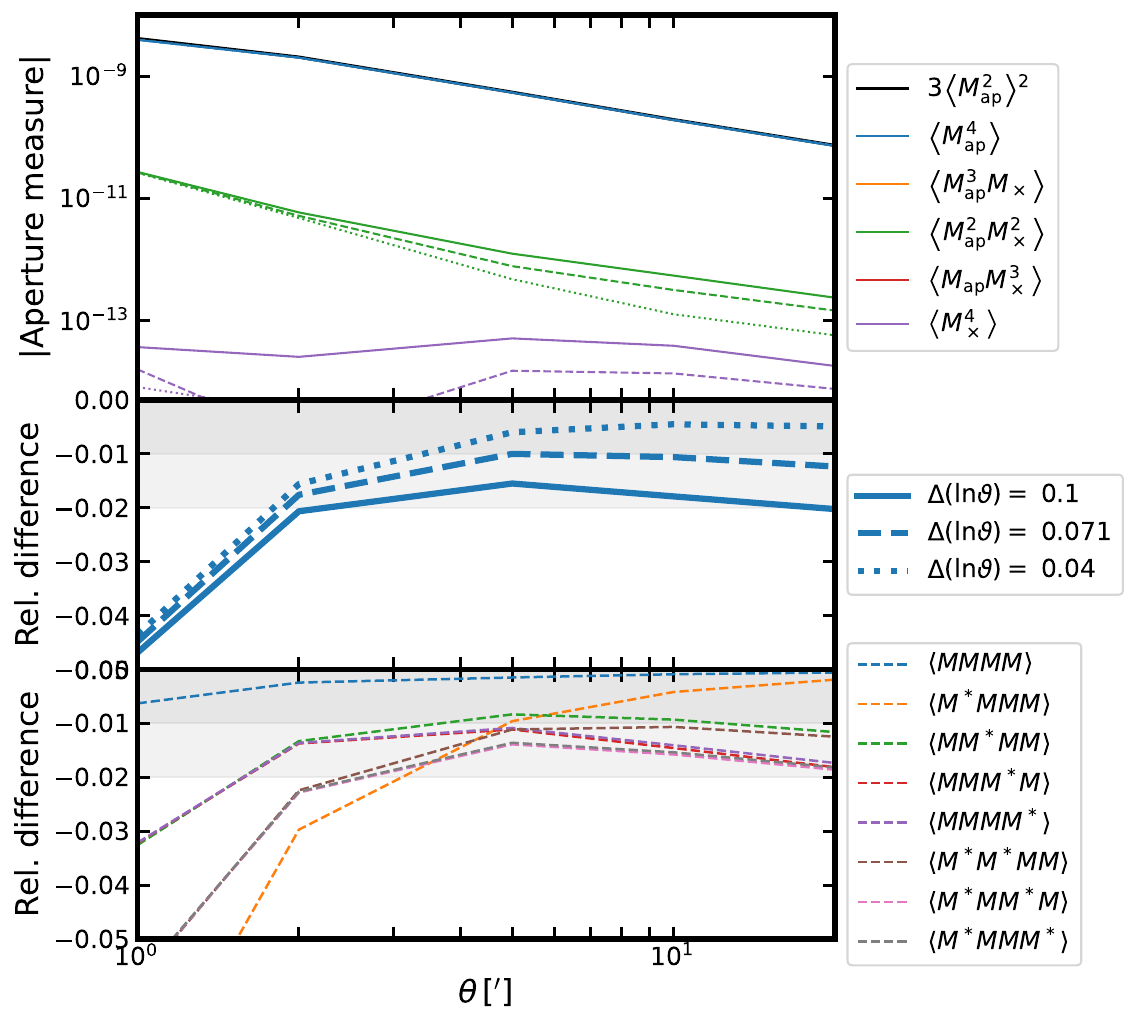}
    \caption{Effect of the binning accuracy on the fourth-order aperture statistics. (Top-middle) Equivalent to \Fig{fig:integration_range} with
    radial integration range $[\m{e}^{-2}\,',\m{e}^5\,']$ and an increasingly finer hypercubical bins, i.e.\ $\Delta_\psi = \Delta(\ln \vartheta)$. (Bottom)  Relative difference of the 8 aperture measures corresponding to the
    8 natural components to the considered ground truth (see text).}
    \label{fig:binning}
\end{figure}

For a given integration range, the apparent convergence
of the integral depends on the binning accuracy. \Fig{fig:binning} shows the convergence of the integrals for an increasing number of radial bins.

At large $\theta$, the dispersion for different natural components arises, for a GRF, from the different weight given to quadrilateral
 configurations at large radii, where the bin size is large. The bottom panel in \Fig{fig:binning} shows that the largest underestimations are found
 for the integrals over $\Ga{\mu}$ with $\mu>1$, as expected from \secref{sec:bin_geometry}.

On scales $\theta \in [2\,', 20\,']$ and radial integration limits $\vt\in[\m{e}^{-2}\,', \m{e}^5\,']$,
87 angular bins and 98 radial bins, i.e.\ $\Delta(\ln \vt) = 0.07$, the obtained results are within a two percent-level precision, with a more precise result at larger $\theta$.

\subsection{Integrated 4PCF}
\label{sec:integrated}
A realistic 4PCF cannot be evaluated at a fixed quadruplet configuration $(\vt_1, \vt_2, \vt_3, \psi_{12}, \psi_{23})$, since the probability of finding a quadruplet
at such exact configuration is formally zero. The 4PCF is then averaged over all quadruplets in a
radial bin, i.e.\ integrated over the radial grid of bin width $\Delta_{\vt;i}$, for the $i$-th bin
\begin{align}
    \overline{\Ga{\mu}}^{\times}_{i,j,k,m,n} =&  \frac{ 1}{\Delta_{\vt;i} \Delta_{\vt;j} \Delta_{\vt;k}}\int_{\vt_{1;i}-0.5\Delta_{\vt;i}}^{\vt_{1;i}+0.5\Delta_{\vt;i}}\diff \vt_1 
    \int_{\vt_{2;j}-0.5\Delta_{\vt;j}}^{\vt_{2;j}+0.5\Delta_{\vt;j}} \diff \vt_2 \nonumber \\
   & \hspace{-1cm}\times\int_{\vt_{3;k}-0.5\Delta_{\vt;k}}^{\vt_{3;k}+0.5\Delta_{\vt;k}} \diff \vt_3\ \Ga{\mu}^\times
    \left(\vt_{1}, \vt_{2}, \vt_{3}, \psi_{12;m}, \psi_{23;n}\right).
\end{align}
By using a multipole-based estimator (described in \citetalias{porth2025}) we are free to choose the angular positions within a precision determined by the number
of multipoles. The estimate of the aperture measure for the $\mu$-th natural component reads
\begin{eqnarray}
    \begin{aligned}
        \left<M^{(*)4}\right>_\m{Rie,int}^{\mu}(\theta) =& \sum_{i, j, k =1}^{N_\m{rad}}\frac{\vt_{i}\Delta_{\vt;i}}{\theta^2} \frac{\vt_{j} \Delta_{\vt;j}}{\theta^2}\frac{\vt_{k} \Delta_{\vt;k}}{\theta^2} \\
        &\times \sum_{m, n = 1}^{N_{\m{ang}}}\frac{\Delta_{\psi;m}}{2\pi} \frac{\Delta_{\psi;n}}{2\pi}\  \overline{\Ga{\mu}}^\times_{i,j,k,m,n}\ K_{\mu;ijkmn},
        \label{eq:integrated}
    \end{aligned}
\end{eqnarray}
which could potentially be computed from a realistic 4PCF. For notational simplicity we defined
\begin{equation}
    K_{\mu;ijkmn} = K_\mu \left(\frac{\q{0;ijkmn}}{\theta},\frac{\q{1;ijkmn}}{\theta},\frac{\q{2;ijkmn}}{\theta},\frac{\q{3;ijkmn}}{\theta}\right),
\end{equation}
with $\bb{q}_s$ from \Eq{eq:CMV_vectors},
$\vt_{1;i}$ the arithmetic centre of the $i$-th radial bin, $\psi_{23;m}$ the arithmetic centre of the $m$-th angular bin, and equivalent for the other radial and angular coordinates.

\subsubsection{Riemann integrated 4PCF}
\label{sec:Riemann_integrated_4pcf}
To emulate the implicit bin-averaging in practical $n$PCF estimators, we estimate the radially integrated 4PCF by subdividing the original radial bins into $r$ finer bins of width $\delta_{\vt;i'}$,
such that $\sum_{i'=1}^{r} \delta_{\vt;i'} = \Delta_{\vt;i}$. The estimate of the $\mu$-th natural component is
\begin{eqnarray}
\begin{aligned}
    &\overline{\Ga{\mu}^{\m{Rie}}}^\times_{i,j,k,m,n} =  \frac{1}{ 
        \sum_{i', j', k' = 1}^{r} \delta_{\vt;i'} \delta_{\vt;j'}
        \delta_{\vt;k'}}\\
        &\times \sum_{i', j', k'=1}^{r} \Ga{\mu}^\times \left(\vt_{i'}, \vt_{j'}, \vt_{k'}, \psi_{12;m}, \psi_{23;n}\right)\delta_{\vt;i'} \delta_{\vt;j'} \delta_{\vt;k'},
    \label{eq:Riemann_sum_shear}
\end{aligned}
\end{eqnarray}
with $\vt_{i'}$ the arithmetic centre of the $i'$-th fine bin within a coarse bin.

\begin{figure}[htbp]
    \centering
    \includegraphics[width=.999\linewidth,valign=t]{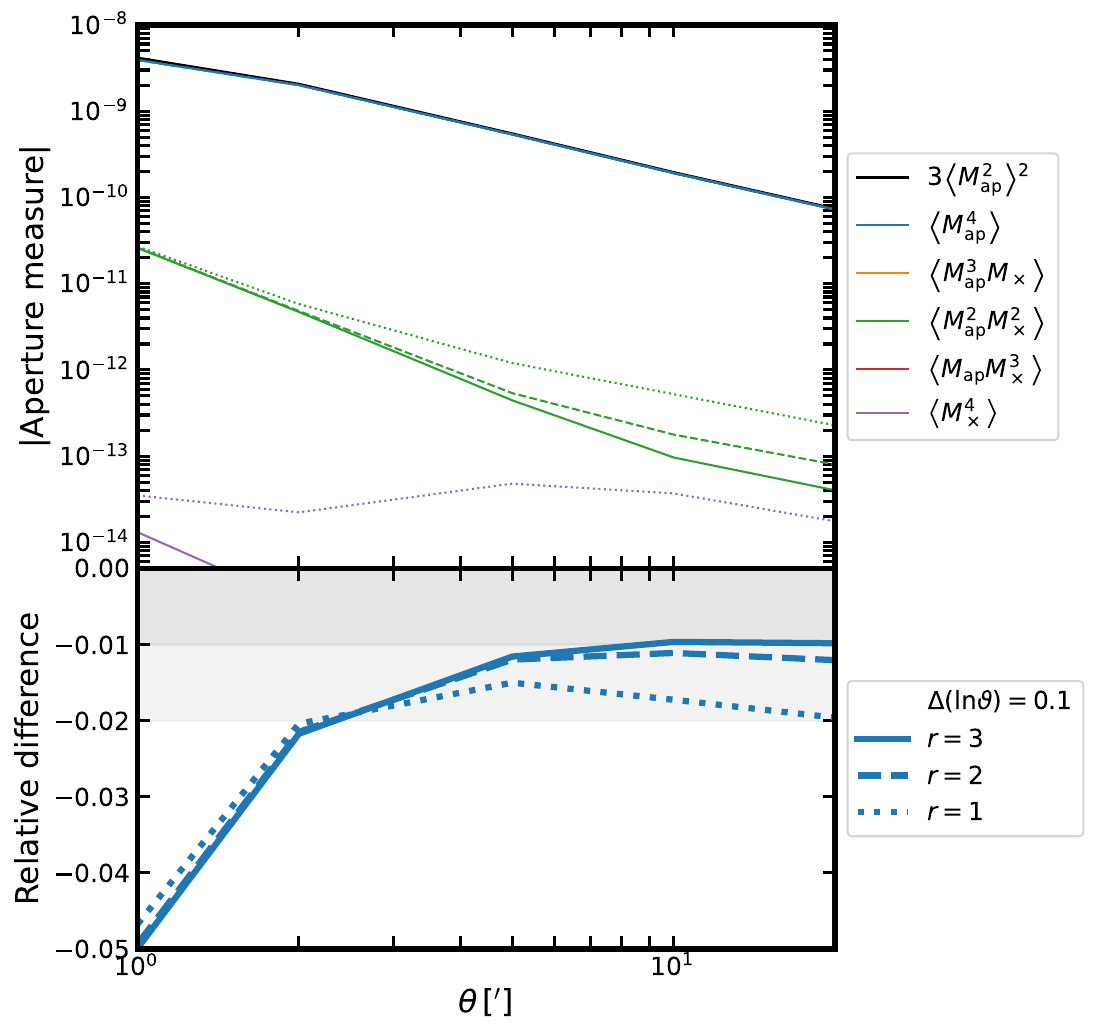}
    \caption{Increase on the precision on the fourth-order aperture statistics by integrating the 4PCF within a bin with a Riemann sum.
    Equivalent to \Fig{fig:integration_range} with increasingly finer bins for the 4PCF, $N_\m{rad,fine}=rN_\m{rad,coarse}$, and $N_\m{rad,coarse}=70$ bins for the filters [$\Delta(\ln \vt) = 0.10$].}
    \label{fig:integrated}
\end{figure}

Figure \ref{fig:integrated} shows $\left<M^{(*)4}\right>_\m{Rie,int}^{p}(\theta)$ from \Eq{eq:integrated} with the average shear estimated
with \Eq{eq:Riemann_sum_shear}. The number of angular bins is such that the fine grid has hypercubical bins.
In all cases we consider $N_\m{rad,coarse}=70$, i.e.\ $\Delta(\ln \vt) = 0.10$, and an increasing $r$.

Integrating the 4PCF within a coarse bin reduces the finite bin effects on large scales, thus reducing the dispersion between the different aperture measures.
Physically, this means that close-by configurations are now accounted for in the integration scheme also at larger scales.
Even with a rather coarse binning of $\Delta(\ln \vt) = 0.10$ for the filter we achieve now a high precision on the aperture statistics.
This is a favorable result towards the application of the fourth-order shear statistics, because an increase in the number of radial bins for the 4PCFs comes at a larger computational cost.

%--------------------------------------------------------------------

\section{Cosmology with second-, third-, and fourth-order statistics}
\label{sec:forecast}
\subsection{Datavector on simulated datasets}
\label{sec:sims}
We consider as observables the equal-scale aperture statistics
\begin{equation}
    \bb{m} = \left\{\left< M_\m{ap}^2\right>(\theta_\alpha),\left< M_\m{ap}^3\right>(\theta_\beta),\left< M_\m{ap}^4\right>_c(\theta_\gamma)\right\},
\end{equation}
with $N_\theta$ aperture radii\footnote{The simulation suite \textit{cosmo}-SLICS \citep{cosmoSLICS2019} becomes inaccurate at scales $\theta\lesssim 4\,'$, where we find the relative error on second-order statistics to be larger than $3\%$ w.r.t. the predictions from the non-linear cold-dark-matter Bacco emulator \citep{nonlinearbacco2021}.} 
\begin{equation}
    %\theta \in\{\ang[padangle=none]{;4.00;},\ang[padangle=none]{;5.66;}, \ang[padangle=none]{;8.00;}, \ang[padangle=none]{;11.31;}, \ang[padangle=none]{;16.00;}\}.
    \theta \in\{\ang{;4.00;},\ang{;5.66;}, \ang{;8.00;}, \ang{;11.31;}, \ang{;16.00;}\}.
\end{equation}
We compute the covariance matrix between the observables, $\tens{C}$, of dimension $(3N_\theta)^2$, from the full-sky ray-tracing simulations by \citeauthor{Takahashi2017} (\citeyear{Takahashi2017}, hereafter T17), assuming a DES-Y3 \citep{desy3} redshift distribution and shape noise; for more details we refer to the internally estimated covariance from Sect. 6.3.1 in \citetalias{porth2025}.

We extract the cosmological dependence of the observables from the \emph{cosmo}-SLICS simulations by \citeauthor{cosmoSLICS2019} (\citeyear{cosmoSLICS2019}, hereafter cS19). cS19 consists of 26 sets of cosmological parameters,
\begin{equation}
    \bb{\mu} = \{\Omegam, S_8, h, w_0\},
    \label{eq:parameters_CS}
\end{equation}
distributed on a latin hypercube. For each cosmology there are 25 light cones, each one with 15 to 28 noiseless convergence maps of $100 \deg^2$ at equidistant comoving distance and $z<3.0$ \citep{cosmoSLICS2019}. We create DES-Y3-like convergence maps assuming the fiducial DES-Y3 redshift distribution and we estimate the aperture mass moments with the direct estimator from \Eq{eq:direct}.\footnote{The aperture mass moments are computed with the direct estimator from \Eq{eq:direct}, using the convolution theorem
\begin{equation}
    M_\m{ap}(\bvt,\theta) = \mathcal{F}\left[\hat{u}(\theta l)\hat{\kappa}(\bb{\ell})\right],
\end{equation}
and Fast Fourier Transforms. In order to avoid finite field effects, we further crop a $4\theta$ boundary on the fields, as suggested by \citet{heydenreich2023}.}

We estimate the derivatives of the observables w.r.t. the cosmological parameters with a simulation-based approach. Since the statistics measured on the cS19 are noisy, an estimate through finite differences becomes unstable. We propose a more robust method, where the tangent plane to the observable is constructed by a loss function minimization, which can be easily generalized to high-dimensional parameter space and can account for irregular sampling around the fiducial cosmology, as we find for example with a latin hypercube simulation suite. In \appref{app:derivatives} we describe the method and test for the stability of the derivatives on cS19.

\subsection{Fisher forecast formalism}
\label{sec:fisher}
The likelihood of the observables given the parameters is
\begin{eqnarray}
\begin{aligned}
    \mathcal{L}(\bb{m},\bb{\mu}) = \m{exp}&\left\{-\frac{1}{2}\sum_{ij}\left[m_i(\bb{\mu})-m_i(\bb{\mu}_\m{fid})\right]\right. \\
    &\times\left. \vphantom{-\frac{1}{2}\sum_{ij}\left[m_i(\bb{\mu})-m_i(\bb{\mu}_\m{fid})\right]} \tens{C}^{-1}_{ij}\left[m_j(\bb{\mu})-m_j(\bb{\mu}_\m{fid})\right] \right\},
    \label{eq:likelihood}
\end{aligned}
\end{eqnarray}
where $\bb{\mu}_\m{fid}$ are the fiducial values of the parameters (here matched to the cosmological set in cS19 labeled as \verb|fid|, hereafter \verb|cSfid|).

An accurate determination of the parameter posteriors is possible with Bayesian inference, see e.g.\ \citet{Burger2024} for the joint analysis of second- and third-order aperture statistics. However, this requires extensive computational power for the exploration of parameter space and accurate theoretical models for fourth-order shear statistics, both of which are not available.

Assuming a Gaussian likelihood of the observables, the maximum likelihood estimate of the parameters corresponds to the maximum a posteriori from Bayesian inference. The uncertainty on the parameters can then be estimated via the expectation value of the Hessian of the log-likelihood, $\ln{\mathcal{L}}$, i.e.\ the Fisher information matrix
\begin{equation}
    \tens{F}_{qp} = \left<\frac{\partial^2\left(-\ln{\mathcal{L}}\right)}{\partial \mu_q \partial \mu_p}\right>.
    \label{eq:fihser_0}
\end{equation}

If the dependence of the covariance matrix of the observables, $\tens{C}$, on the cosmological parameters, $\bb{\mu}$, can be neglected, then
\begin{equation}
    \tens{F}_{qp} = \sum_{i,j} \left.\frac{\partial m_i}{\partial \mu_q}\right|_{\bb{\mu}_\m{fid}} \tens{C}_{ij}^{-1} \left.\frac{\partial m_j}{\partial \mu_p}\right|_{\bb{\mu}_\m{fid}}
    \label{eq:fisher_cov}
\end{equation}
can be estimated through the covariance at the fiducial value (here from T17) and the derivatives at the fiducial value.

Moreover, if the observables are unbiased estimators of their underlying distributions, then $\sqrt{\tens{F}^{-1}_{qq}} = \sigma_q$ is the best possible precision on the cosmological parameters with the given observables, according to the Cram\'er-Rao theorem \citep{rao_fisher, cramer_fisher}, and $\tens{F}^{-1}$ is the covariance matrix between them.

%\subsection{Estimation of derivatives}

%We estimate the derivatives from \Eq{eq:fisher_cov} with a simulation-based approach. Since the statistics measured on the cS19 are noisy, an estimate through finite differences becomes unstable. We propose a more robust method, where the tangent plane to the observable is constructed by a loss function minimization, which can be easily generalized to high-dimensional parameter space and can account for irregular sampling around the fiducial cosmology, as we find for example with a latin hypercube simulation suite. 

%In \appref{app:method} we describe the method for some general observables and parameters. In \appref{app:derivatives_cS} we specify it for the cS19 simulations and study the stability of the derivatives for the aperture statistics. We find that derivatives are stable when considering the cosmological parameters $\Omegam$ and $\sigma_8$ and marginalizing over $h$ and $w_0$, but become unstable if we try to constrain all four parameters simultaneously. When using sparse simulations around the fiducial value, as cS19 around \verb|cSfid|, the derivatives are biased towards lower values, resulting in overestimated Fisher ellipses for the independent aperture statistics, and possibly under- or overestimated for the joint analysis (see \Fig{fig:derivatives}). Even on such sparse simulations, we recover a value of the mean for second- and third-order statistics compatible within their $1\sigma$ precision to the measured one.

\subsection{Combined forecast second- and higher-order aperture statistics}
\label{sec:forecastfourth}
\begin{SCfigure*}[][htbp]
        \centering
        \includegraphics[width=.6\textwidth]{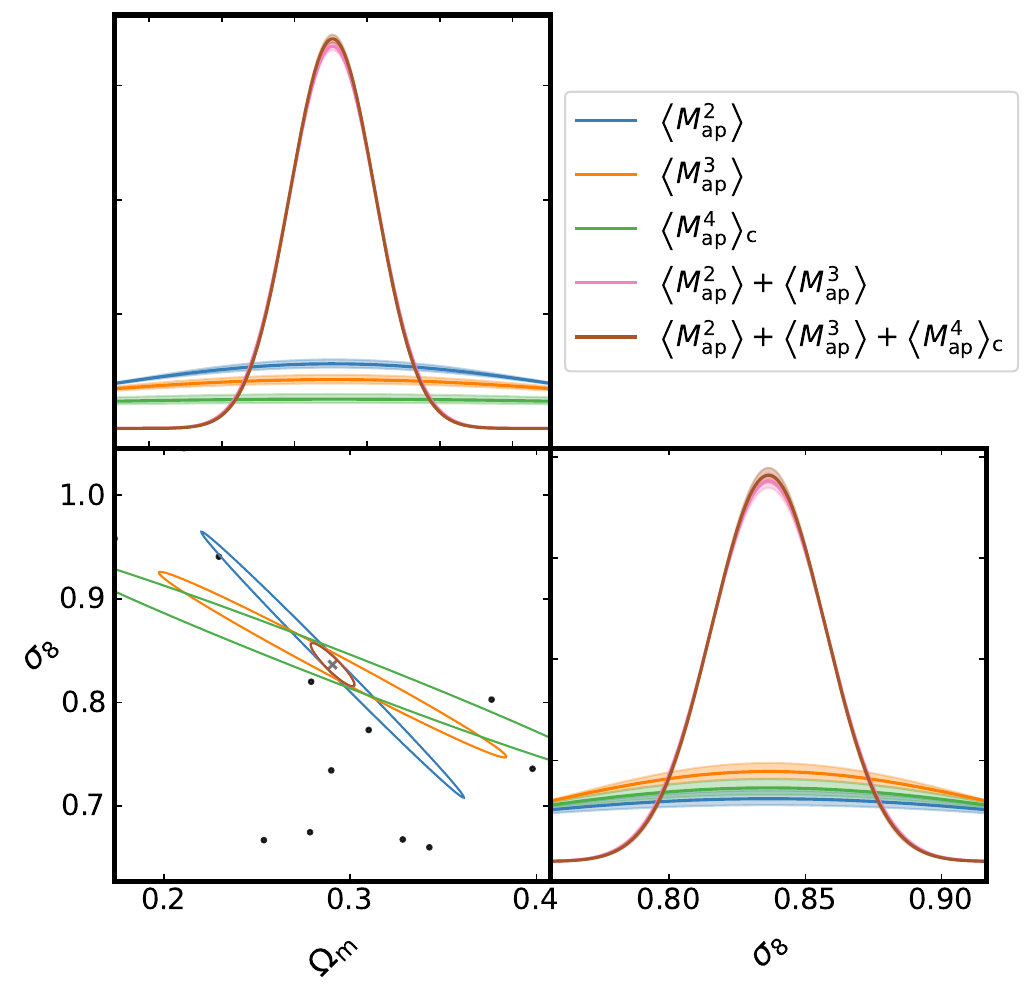}
        \cprotect\caption{Joint forecast of second-, third-, and fourth-order aperture statistics, with 1$\sigma$ contours from the Fisher covariance matrix $\tens{F}^{-1}$. The cloud of points in the two-dimensional distribution corresponds to the closest sets of cosmological parameters to the fiducial \verb|cSfid| value, marked with a cross.}
        \label{fig:forecast}
\end{SCfigure*}

In \Fig{fig:forecast} we show the 1$\sigma$ Fisher ellipse for $\{\Omegam,\sigma_8\}$ for $\left<M_\m{ap}^2\right>$, $\left<M_\m{ap}^3\right>$, and $\left<M_\m{ap}^4\right>$ individually, the joint analysis of second- and third-order statistics, and the joint analysis of second-, third-, and fourth-order statistics, along with the one-dimensional distributions (PDF). 

The shaded regions in the PDF represent the $1\sigma$ Jackknifing error bars over the cS19 light cones, which affect the width of the Gaussian PDF but not the fiducial \verb|cSfid| value. This error arises from the reduced number of available simulations and does not account for other sources of uncertainty, such as cosmic variance on scales larger than $10\degree$ or shape noise. We perform a similar analysis on the two-dimensional distributions in \appref{app:stability_ellipses}. 

%Upon comparison with the Fisher ellipses from theory, we find that the constraining power of a joint second- and third-order statistics is overestimated in \Fig{fig:forecast} compared to the individual orders; i.e.\ there are some artificial degeneracy breaking to third-order, when using spare simulations.

In our setting and within the set of assumptions we cannot report a significant improvement in the constraining power of the aperture statistics when including fourth-order statistics. If there is any information content on the non-tomographic equal-scale fourth-order aperture statistics, we would require the precision of Stage IV surveys, like $\textit{Euclid}$ \citep{euclidoverview} and a larger set of simulations or an accurate theoretical model in order to retrieve it.

%--------------------------------------------------------------------
\section{Discussion}
\label{sec:discussion}
The statistical power of Stage IV surveys, like \textit{Euclid} \citep{euclidoverview}, opens a new range of possibilities for higher-order statistics. Higher moments of the shear field, in form of shear $n$-point correlation functions or their projections into the $n$-th order aperture statistics, are expected to contain part of the non-Gaussian information of the field in their connected part.

Here we explored the relation between $n$-th order aperture statistics and shear $n$-point correlation functions ($n$PCF) or $n$-point polyspectra. In particular, we derived the explicit expressions for the filters of the fourth-order aperture statistics in terms of the 4PCF and in terms of the trispectrum. On an observed field, the computation through the 4PCF is necessary, as it is not biased by masked regions in the observational field.

We tested the filters for the 4PCF on a model for a shear Gaussian random field (GRF), where the fourth-order statistics only contain disconnected terms, i.e.\ can be expressed in terms of the second-order statistics. We proposed and tested an integration scheme that provides a two-percent-level precision on the aperture statistics for a GRF, which remains well below noise budget for Stage IV surveys.

We find deviations from the expected values on very small and very large scales due to an $E$- and $B$-mode mixing, similar as the one described in \citet{kilbinger2006} for second-order aperture statistics. This mode-mixing enters fourth-order statistics through an incomplete radial integration range (see \Fig{fig:integration_range}) and an insufficient binning accuracy (see \Fig{fig:binning}). Despite the stronger impact of these effects on some of the complex aperture measures (see \Fig{fig:radial_integration}), a separation into modes is only possible 
%for a general field, with tomography and multi-scale aperture statistics, 
when all the complex aperture measures are considered. We achieve the desired precision for an observational range of $\theta \in [2\,',20\,']$, a radial integration range of $\vt\in[\m{e}^{-2}\,',\m{e}^5\,']$, and a logarithmic binning accuracy of $\Delta(\ln \vt) = 0.07$ with hypercubical bins.

The $E$- and $B$-mode mixing should be taken into account when obtaining the connected fourth-order statistics for a general field, as the predictions from second-order statistics only correspond to the disconnected terms of $\left<M_\m{ap}^4\right>$ up to a degree determined by the integration range and binning accuracy. 

Moreover, we find that using a more realistic model for the 4PCF, where the shear is integrated over radial bins, reduces the required binning accuracy on the 4PCF to $\Delta(\ln \vt) = 0.10$, which is broad enough to be estimated on stage III data. \citetalias{porth2025} develop a multipole-based 4PCF estimator and use the integration method described here to obtain a significant detection of fourth-order aperture statistics. They also find that the error estimations presented here for a GRF closely match the observed errors in the $\langle M_\m{ap}^4\rangle_\m{c}$-statistics measured on DES-Y3 \citep{desy3}.
%, meaning that most of the uncertainty to fourth-order is carried on the disconnected part.

Finally, we computed a Fisher forecast for joint second-, third-, and fourth-order aperture statistics on a DES-Y3-like set-up. To do so, we suggest a robust method for estimating derivatives from simulations, which we test on the \textit{cosmo}-SLICS \citep{cosmoSLICS2019} simulation suite.

We do not find a significant improvement in the constraining power of the aperture statistics when including fourth-order statistics to a joint $\langle M_\m{ap}^2\rangle+\langle M_\m{ap}^3\rangle$ joint analysis, which we believe is driven by the given set of assumptions. These assumptions are:
\begin{itemize}
    \item Due to the nature of Fisher forecasts, we assume a Gaussian distribution of the observables, which is not a good approximation for $\left<M_\m{ap}^3\right>$ \citep{higher_order_euclid} nor $\left<M_\m{ap}^4\right>_\m{c}$ \citepalias{porth2025}. The Fisher ellipses are thus not able to fully capture the degeneracies of the observables.
    \item From a theoretical point of view, we assumed that equal-scale aperture statistics contain most of the information to the given order, if a sufficiently large part of the field is available. This is exact for $\left<M_\m{ap}^2\right>$ and a good approximation for $\left<M_\m{ap}^3\right>$ if one considers tomography (see e.g. \citealp{Burger2024}). Equal-scale aperture moments contain, for increasing order, a smaller fraction of the total information to the given order. Due to the strong correlation between neighboring scales and neighboring orders seen in the covariance matrix from T17 (see Fig. 4 in \citetalias{porth2025}), we expect that a few combinations of unequal scales contain most of the information on the multi-scale aperture moments, and leave this analysis to future work.
    \item From the parameters considered in cS19, we marginalized over the dependence on $h$ and $w_0$. This reduces the relative constraining power of higher-order aperture statistics w.r.t. second-order ones, if their dependence on such parameters is stronger than for second-order statistics.\\
    We recovered a low constraining power on the evolution parameters, such as $w_0$, due to lack of tomography. When studying the derivatives from the samples in \appref{app:derivatives_cS} we find a small scatter in the derivatives with respect to $\Omegam$ that is reduced if we also consider the parameter $w_0$, i.e. the scatter on $\Omegam$ is influenced by the given dark energy model. Considering tomography would weaken this degeneracy due to the different evolution of the different energy components.
    \item Higher-order aperture mass moments have an increasingly stronger decay with aperture radius, so most of the information in fourth-order statistic is at small scales (assuming no shape noise), which are not available in the simulations used for the derivatives, and are even more challenging to model. 
    \item Furthermore, we studied the effect of using the derivatives from theory on the Fisher ellipses for second-order, third-order, and the joint second- and third-order statistics. We find that the relative constraining power of a joint analysis $\left<M_\m{ap}^2\right>+\left<M_\m{ap}^3\right>$ is overestimated when computing derivatives from cS19. This unrealistic degeneracy breaking for a joint second- and third-order analysis can result in a lower relative constraining power of fourth-order statistics.\footnote{We compute the relative constraining power as the ratio of the correspondent Figures of Merit, see \Eq{eq:FoM}.}
\end{itemize}

In our analysis we do not consider any nuisance parameters. Even if fourth-order statistics would be fully degenerate with the lower-order moments (after relaxing all mentioned assumptions), they will have a different degeneracy direction for the nuisance parameters. \citet{pyne2021} suggest the use of third-order statistics to calibrate second-order statistics. On Stage IV surveys, where both second- and third-order moments are used to constrain cosmology, fourth-order statistics would become a consistency check and a test for systematics.

We therefore conclude that performing a full cosmological analysis for fourth-order aperture statistics on a Stage III survey, like DES-Y3 \citep{desy3}, would lead only to a minimal improvement on the cosmological constraints.

We developed and thoroughly tested a method targeted to Stage IV surveys, like \textit{Euclid} \citep{euclidoverview}, where the unprecedented precision will increase the significance of higher-order statistics, allowing for example for a tomographic analysis of $\left<\mapfour\right>_\m{c}$. Under that setting, it will be possible to determine whether fourth-order shear statistics break degeneracies between second- and third-order statistics, or if they are degenerate with the previous orders, in which case they can be used to help in the determination of the nuisance parameters.

%--------------------------------------------------------------------

\begin{acknowledgements}
We would like to thank Joachim Harnois-D\'eraps for providing us access to the \textit{cosmo}-SLICS mock data. LP acknowledges support from the DLR grant 50QE2302. LL is supported by the Austrian Science Fund (FWF) [ESP 357-N].
\end{acknowledgements}

\bibliographystyle{aa} % style aa.bst
\bibliography{lit} % your references Yourfile.bib

\begin{appendix}
\onecolumn
\section{Conversion from $n$PCF to aperture statistics}\label{app:NthOrderMap4}

\subsection{Derivation of filters}
\label{app:NthOrderMap4Same}
In order to obtain a $E$/$B$-decomposed aperture statistics it is necessary to convert the $2^{n-1}$ natural components of the shear \npcf \ to their corresponding complex aperture measures. We only present a detailed derivation for the special case of all aperture radii being equal, $\theta_1 = \theta_2 = \cdots, \theta_{n-1} \equiv \theta$ and later mention the modifications to the expressions when allowing for arbitrary configurations of aperture radii. 

As the derivations for all filter functions, $F^{(n)}_\mu$, are structurally equivalent, we only explicitly derive the filter $\left\langle (M^*)^p M^{n-p}\right\rangle$, corresponding to the $\mu_p$th natural component of the \npcf. 
We begin by reordering the expression for the $\left\langle (M^*)^p M^{n-p}\right\rangle$ correlator:
\begin{align}\label{eq:MpMstarNminnusp4IntegralStart}
    &\left\langle (M^*)^p M^{n-p} \right\rangle(\theta)
    \nonumber \\ &\hspace{1cm}\equiv
    \int \dd^2 \mathbf{X}_0 \int \dd^2 \bvartheta_1  \cdots  \int \dd^2 \bvartheta_{n-1} \ Q_\theta(|\mathbf{X}_0|) \cdots Q_\theta(|\mathbf{X}_0+\bvartheta_{n-1}|) \, 
    \nonumber \\ &\hspace{2cm} \times \ \ 
    \left\langle \g_{\mathrm{cart}}^*(\mathbf{X}_0) \cdots \g_{\mathrm{cart}}^*(\mathbf{X}_0+\bvartheta_{p-1}) \g_{\mathrm{cart}}(\mathbf{X}_0+\bvartheta_p) \cdots \g_{\mathrm{cart}}(\mathbf{X}_0+\bvartheta_{n-1}) \right\rangle \ \ee^{-2\ii \left(-\zeta_{0} - \cdots -  \zeta_{p-1} +  \zeta_{p} + \cdots +  \zeta_{n-1}\right)}
    \nonumber \\ &\hspace{1cm}=
    \int \dd^2 \bvartheta_1  \cdots  \int \dd^2 \bvartheta_{n-1} \ \Gamma_{\mu_p}^{\rm cart}(\bvartheta_1, \cdots, \bvartheta_{n-1}) \int \dd^2 \mathbf{X}_0 \ \frac{\left(\breve{\mathbf{X}}_0\right)^2}{\left|\breve{\mathbf{X}}_0\right|^2} \, \cdots \, \frac{\left(\breve{\mathbf{X}}_0+\breve{\bvartheta}_{p-1}\right)^2}{\left|\breve{\mathbf{X}}_0+\breve{\bvartheta}_{p-1}\right|^2} \frac{\left(\breve{\mathbf{X}}^*_0+\breve{\bvartheta}^*_{p}\right)^2}{\left|\breve{\mathbf{X}}_0+\breve{\bvartheta}_{p}\right|^2} \, \cdots \, \frac{\left(\breve{\mathbf{X}}^*_0+\breve{\bvartheta}^*_{n-1}\right)^2}{\left|\breve{\mathbf{X}}_0+\breve{\bvartheta}_{n-1}\right|^2}
    \nonumber \\ &\hspace{2cm} \times \ \ 
    \frac{\left|\breve{\mathbf{X}}_0\right|^2 \cdots \left|\breve{\mathbf{X}}_0+\breve{\bvartheta}_{n-1}\right|^2}{(4\pi \theta^4)^n} \ \ \exp{-\frac{1}{2\theta^2}\left( \left|\breve{\mathbf{X}}_0\right|^2 + \sum_{\ell=1}^{n-1} \left|\breve{\mathbf{X}}_0+\breve{\bvartheta}_{\ell}\right|^2 \right)}
    \nonumber \\ &\hspace{1cm}=
    \int \dd^2 \bvartheta_1  \cdots  \int \dd^2 \bvartheta_{n-1} \ \Gamma_{\mu_p}^{\rm cart}(\bvartheta_1, \cdots, \bvartheta_{n-1}) \ \ \exp{-\frac{1}{2\theta^2}\sum_{\ell=0}^{n-1}\left|\mathitq_\ell\right|^2}
    \nonumber \\ &\hspace{2cm} \times \ \ 
    \int \dd^2 \mathbf{X}_0 \ \frac{\left(\breve{\mathbf{X}}_0\right)^2 \cdots \left(\breve{\mathbf{X}}_0+\breve{\bvartheta}_{p-1}\right)^2\left(\breve{\mathbf{X}}^*_0+\breve{\bvartheta}^*_{p}\right)^2 \cdots \left(\breve{\mathbf{X}}^*_0+\breve{\bvartheta}^*_{n-1}\right)^2}{(4\pi \theta^4)^n} \ \ \exp{-\frac{n}{2\theta^2}\left|\mathbf{X}_0-\mathitq_0\right|^2} \ \ ,
\end{align}
where in the second step we used the expression \eqref{eq:filter_crit} of the $Q$-filter and in the third step rewrote the exponential using the $n$ vectors $\{\mathitq_\mu\}$ connecting the centroid of the points $\{\mathbf{X}_\mu\}$ with $\mathbf{X}_\mu$; $\mathitq_0 = -\frac{1}{n}\sum_{i=1}^{n-1} \bvartheta_i$ and $\mathitq_i = \mathitq_0 + \bvartheta_{i}$; and further used the property $\sum_{\mu=0}^{n-1}\mathitq_\mu = 0$.

Now focusing our attention to the innermost integral, $I$, we can make the substitution $\mathbf{Z} = \mathbf{X}_0 - \mathitq_0$ and rewrite $I$ in polar coordinates:
\begin{align}\label{eq:MapnHelper1}
    I &\sim
    \int \dd Z \ Z \ \ee^{-\frac{n}{2\theta^2}Z^2} \int_0^{2\pi} \dd \varphi_Z \ 
    \left[\prod_{\ell=0}^{p-1} \left(\breve{Z} + \breve{\mathitq}_{\ell} \right)^2 \right] \, \left[\prod_{\ell=p}^{n-1} \left(\breve{Z}^* + \breve{\mathitq}_{\ell}^* \right)^2 \right] \ .
\end{align}
In this form we note that when expanding the products, the only terms that give a non-vanishing contribution to $I$ are of the form $c(\breve{\mathitq}_0,\cdots,\breve{\mathitq}_{n-1}) \ \left|\mathbf{Z}\right|^{2k}$. For being able to write down the corresponding prefactors, we make use of Vietas formulas, which give the coefficients when expanding out a factorized polynomial of order $n$,
\begin{align}
    \prod_{i=1}^n \left( x + x_i \right) = \sum_{k=0}^n e_k(x_1, \cdots, x_n) \, x^{n-k} 
    \hspace{.7cm} ; \hspace{.7cm}
     e_k(x_1, \cdots, x_n) = \sum_{1\leq i_1 < \cdots < i_k \leq n }x_{i_1} \cdots x_{i_n} \ \m{and}\ e_0(x_1, \cdots, x_n) = 1 \ ,
\end{align}
where $e_k(x_1, \cdots, x_n)$ denotes the elementary symmetric polynomial of degree $k$ in $n$ variables. We can apply these relations to both products in \eqref{eq:MapnHelper1}, evaluate the integral over $\psi_1$ order-by-order and solve the remaining Gaussian integral, yielding:
\begin{align}\label{eq:IintegralOrderNSolution}
    I &\sim 
    %\int \dd \widetilde{X}_1 \ \widetilde{X}_1 \ \ee^{-\frac{N}{2\theta^2}\widetilde{X}_1^2} \int_0^{2\pi} \dd \psi_1 \ \left[\sum_{\ell=0}^{2p} e_{\ell}(\mathitq_1,\mathitq_1,\mathitq_2,\mathitq_2,\cdots,\mathitq_p,\mathitq_p) \ \left(\widetilde{\mathbf{X}}_1\right)^{2p-\ell}\right] \, 
    %\nonumber \\ &\hspace{2cm}\times 
    %\left[\sum_{\ell'=0}^{2(N-p)} e_{\ell'}(\mathitq^*_{p+1},\mathitq^*_{p+1},\mathitq^*_{p+2},\mathitq^*_{p+2},\cdots,\mathitq^*_N,\mathitq^*_N) \ \left(\widetilde{\mathbf{X}}_1^*\right)^{2(N-\ell)-\ell'}\right]
    %\nonumber \\ &=
    %(2\pi) \ \int \dd \widetilde{X}_1 \ \widetilde{X}_1 \ \ee^{-\frac{N}{2\theta^2}\widetilde{X}_1^2} \left[ \sum_{k=0}^{2p} \left(\widetilde{X}_1\right)^{2k} \ e_{2p-k}(\mathitq_1,\mathitq_1,\mathitq_2,\mathitq_2,\cdots,\mathitq_p,\mathitq_p) \ \  e_{2(N-p)-k}(\mathitq^*_{p+1},\mathitq^*_{p+1},\mathitq^*_{p+2},\mathitq^*_{p+2},\cdots,\mathitq^*_N,\mathitq^*_N) \right] 
    %\nonumber \\ &=
    (2\pi) \sum_{k=0}^{2p} \frac{k! \, 2^k \, \theta^{2(k+1)}}{n^{k+1}} \ e_{2p-k}(\breve{\mathitq}_0,\breve{\mathitq}_0,\breve{\mathitq}_1,\breve{\mathitq}_1,\cdots,\breve{\mathitq}_{p-1},\breve{\mathitq}_{p-1}) e_{2(n-p)-k}(\breve{\mathitq}^*_{p},\breve{\mathitq}^*_{p},\breve{\mathitq}^*_{p+1},\breve{\mathitq}^*_{p+1},\cdots,\breve{\mathitq}^*_{n-1},\breve{\mathitq}^*_{n-1}) \ .
\end{align}
%where in the first step we substituted Vietas formulas, in the second step only kept the terms that will survive the $\psi_1$ integration and in the final step solved the Gaussian integral. 
We note that the $e_k$ themselves can readily be allocated recursively via the Newton identities. Finally, plugging \eqref{eq:IintegralOrderNSolution} into \eqref{eq:MpMstarNminnusp4IntegralStart}, substituting the relation between the $\times$-projection and the cartesian projection of $\Gamma_{\mu_p}$ and rearranging a bit we have
\begin{align}
    &\left\langle (M^*)^p M^{n-p} \right\rangle(\theta) = 
    \int \frac{\dd \vartheta_1 \,  \vartheta_{1}}{\theta^2} \, \cdots \, \int \frac{\dd \vartheta_{n-1} \, \vartheta_{n-1}}{\theta^2} \, \int \frac{\dd \phi_{1\,2}}{2\pi} \cdots \int \frac{\dd \phi_{1\,n-1}}{2\pi} 
    \nonumber \\ &\hspace{2cm} \times
     \ \Gamma_{\mu_p}^{\times}(\bvartheta_1, \cdots, \bvartheta_{n-1}) \ \ee^{2\ii\left[\sum_{\ell=2}^{n-2}\beta^{\mu_p}_{\ell+1}\phi_{1\,\ell}+\left(0.5\beta^{\mu_p}_{1}+\beta^{\mu_p}_{n}\right)\phi_{1\,n-1}\right]} \ F^{(n)}_{\mu_p}\left(\frac{\breve{\mathitq}_0}{\theta}, \cdots \frac{\breve{\mathitq}_{n-1}}{\theta}\right)
    \ ,
\end{align}
\begin{align}
    F^{(n)}_{\mu_p}(\breve{\mathitq}_0, &\cdots \breve{\mathitq}_{n-1}) 
    = 
    \ee^{-\frac{1}{2}\left( \sum_{\mu=0}^{n-1}|\breve{\mathitq}_\mu|^2\right)} 
    %\ \frac{(\cqac)^2 \cdots (\breve{\mathitq}_p^*)^2 \ \breve{\mathitq}_{p+1}^2 \cdots \breve{\mathitq}_{n}^2 }{|\cqa|^2 \cdots  |\breve{\mathitq}_n|^2}
    %\nonumber \\ &\times 
    \sum_{k=0}^{2p} \frac{k! \  e_{2p-k}(\breve{\mathitq}_0,\breve{\mathitq}_0,\breve{\mathitq}_1,\breve{\mathitq}_1,\cdots,\breve{\mathitq}_{p-1},\breve{\mathitq}_{p-1}) \, e_{2(n-p)-k}(\breve{\mathitq}^*_{p},\breve{\mathitq}^*_{p},\breve{\mathitq}^*_{p+1},\breve{\mathitq}^*_{p+1},\cdots,\breve{\mathitq}^*_{n-1},\breve{\mathitq}^*_{n-1})}{2^{n-k}\,n^{k+1}}
     \ , 
\end{align}
where we performed the integral over $\varphi_1$ owing to statistical isotropy and set $\varphi_1\equiv 0$ in all relations depending on $\varphi_1$.

\subsection{Modifications for unequal aperture radii}
\label{app:NthOrderMap4Diff}
Much of the above derivation can be reused when allowing for different filter radii $\theta_{\mu}$. First, in resemblance to \eqref{eq:MpMstarNminnusp4IntegralStart}, we have
\begin{align}\label{eq:MultiscaleMpMstarNminnusp4IntegralStart}
    \left\langle (M^*)^p M^{n-p} \right\rangle(\theta_0, \cdots, \theta_{n-1}) 
    &=
    \int \dd^2 \bvartheta_1  \cdots  \int \dd^2 \bvartheta_{n-1} \ \Gamma_{\mu_p}^{\rm cart}(\bvartheta_1, \cdots, \bvartheta_{n-1}) \ \ \ee^{-c/2}
    \nonumber \\ &\hspace{-2cm} \times \ \ 
    \int \dd^2 \mathbf{X}_0 \ \frac{\left(\breve{\mathbf{X}}_0\right)^2 \cdots \left(\breve{\mathbf{X}}_0+\breve{\bvartheta}_{p-1}\right)^2\left(\breve{\mathbf{X}}^*_0+\breve{\bvartheta}^*_{p}\right)^2 \cdots \left(\breve{\mathbf{X}}^*_0+\breve{\bvartheta}^*_{n-1}\right)^2}{(4\pi)^n (\theta_0\cdots \theta_{n-1})^4} \ \ \ee^{-\left(\mathbf{X}_0-\mathbf{b}\right)^2/(2a^2)} \ \ ,
    \nonumber \\ &\hspace{1cm}\mathrm{where} \ \
     a^{-2} \equiv \sum_{i=0}^{n-1}\theta_i^{-2}  \ \ , \hspace{1cm} \mathbf{b} \equiv - \, a^2  \sum_{i=0}^{n-1}\frac{\bvartheta_i}{\theta_i^2} \ \ , \hspace{1cm} c \equiv \left(\sum_{i=0}^{n-1}\frac{|\vartheta_i|^2}{\theta_i^2}\right)-\frac{|\mathbf{b}|^2}{a^2} \ .
\end{align}
The integral $I$ is defined and evaluated using the same procedure as in the steps leading to \eqref{eq:IintegralOrderNSolution} and we find
\begin{align}
    I &\sim (2\pi) \sum_{k=0}^{2p} k! \, 2^k \, a^{2(k+1)} \ e_{2p-k}(\breve{\textbf{d}}_0,\breve{\textbf{d}}_0,\breve{\textbf{d}}_1,\breve{\textbf{d}}_1,\cdots,\breve{\textbf{d}}_{p-1},\breve{\textbf{d}}_{p-1}) \, e_{2(n-p)-k}(\breve{\textbf{d}}^*_{p},\breve{\textbf{d}}^*_{p},\breve{\textbf{d}}^*_{p+1},\breve{\textbf{d}}^*_{p+1},\cdots,\breve{\textbf{d}}^*_{n-1},\breve{\textbf{d}}^*_{n-1}) \ ,
\end{align}
where we defined $\breve{\textbf{d}}_k \equiv \breve{\textbf{b}} + \breve{\bvartheta}_k$ and we set $\breve{\bvartheta}_0 \equiv 0$.

\subsection{Separation in $E$- and $B$-modes}
\label{app:NthOrderSeparation}
The $2^{n-1}$ complex aperture measures corresponding to the integral over the natural components contain both $E$- and $B$-mode contributions. The general mode-separating aperture statistic of $n$-th order is
\begin{equation}
    \left\langle \map^{n-q} \mperp^{q} \right\rangle(\theta) = \left\langle \Re{M}^{n-q} \Im{M}^q \right\rangle(\theta)= \left\langle \left(|M|\Re{\eiphi{}}\right)^{n-q} \left(|M|\Im{\eiphi{}}\right)^q \right\rangle(\theta)= \left\langle |M|^n \Re{\eiphi{}}^{n-q} \Im{\eiphi{}}^q \right\rangle(\theta),
\end{equation} 
where $q \in \{0,...,n\}$ is the number of cross apertures and we used the polar notation for the complex aperture $M=|M|\eiphi{}$ and the definition of the aperture mass \Eq{eq:m}. In a similar way, the complex aperture measures 
\begin{equation}
    \left\langle (M^*)^p M^{n-p} \right\rangle(\theta) = \left\langle (|M|\eiphi{-})^p (|M|\eiphi{})^{n-p} \right\rangle(\theta) = \left\langle |M|^n\eiphi{(n-2p)} \right\rangle(\theta).
    \label{eq:mapn_eiphi}
\end{equation}
The integrals over the different natural components with $p$ conjugated shears contain different information about the field and should all be computed in order to extract the maximum information about the field, in the presence of tomography and shape noise. We define $A_{np}$ as the average of the $\binom{n}{p}$ complex aperture measures with $p$ complex apertures, which formally follow the same expression as \Eq{eq:mapn_eiphi}. 

We can express the aperture statistics as a linear combination over the different complex aperture measures with real coefficients $C_{qp}$
\begin{equation}
    \left\langle \map^{n-q} \mperp^{q} \right\rangle(\theta) = \sum_{p=0}^{\lfloor n/2 \rfloor} C_{qp} \times\left( \Re{A_{np}}\ \m{or}\ \Im{A_{np}}\right) ,
    \label{eq:linear_general}
\end{equation}
where we find the real value for $q$ even, and the imaginary one for $q$ odd.

Using the properties of the complex numbers, $\Re{z} = (z+z^*)/2$ and $\Im{z} = (z-z^*)/(2 \ \ii)$, and the commutativity of the ensemble averages with additions, we obtain the coefficients $C_{qp}$ through
\begin{equation}
    \frac{1}{2^n\  \m{i}^q}\left(\eiphi{}+\eiphi{-}\right)^{n-q}\left(\eiphi{}-\eiphi{-}\right)^{q} = \frac{1}{2\  \m{i}^{q_\mathrm{mod2}}}\sum_{p=0}^{\lfloor n/2 \rfloor} C_{qp} \times\left( \eiphi{(n-2p)}+(-1)^q\eiphi{-(n-2p)}\right). 
    \label{eq:separation_coeff}
\end{equation}

The different coefficients can be obtained by comparison of both sides on \Eq{eq:separation_coeff}; see Table \ref{tab:cqp} for the values for $n={3,4,5}$. The coefficients for the pure $E$-mode, $q=0$, further simplify to
\begin{equation}
    C_{0p} = \frac{1}{2^{n-1}}\binom{p}{n}\left[1-\frac{\delta^\m{K}_{p-n/2}}{2}\right].
\end{equation}

\begin{table*}[t] 
\centering
\captionsetup{}
\caption{Coefficients for the linear combinations of complex aperture measures to give an $E$- and $B$-mode separation, as in \Eq{eq:linear_general}. Computed from \Eq{eq:separation_coeff}. The coefficients are only defined for $q\leq n$, being the ones for $n=3$ and $n=4$ compatible with the ones presented by \citet{jarvis2004}. The values marked with $\times$ are undefined by \Eq{eq:separation_coeff}, so we can set them to 0, as in \secref{sec:map4_from_m4}.}
\begin{tabular}{|c|c|c|c|c|c|c|}
\hline
    $n$ & $C_{0p}$ & $C_{1p}$ & $C_{2p}$ & $C_{3p}$& $C_{4p}$ & $C_{5p}$  \\
 \hline 
 3 & $\{1/4,3/4\}$ & $\{1/4,1/4\}$ & $\{-1/4,1/4\}$ & $\{-1/4,3/4\}$& -- & -- \\
 4 & $\{1/8,1/2,3/8\}$ & $\{1/8,1/4,\times \}$ & $\{-1/8,0,1/8\}$ & $\{-1/8,1/4, \times\}$& $\{1/8,-1/2, 3/8\}$ & -- \\
  5 & $\{1/16,5/16,5/8\}$ & $\{1/16,3/16,1/8\}$ & $\{-1/16,-1/16,1/8\}$ & $\{-1/16, 1/16, 1/8\}$& $\{1/16, -3/16, 1/8\}$ & $\{1/16, -5/16, 5/8\}$\\
 \hline
\end{tabular}
\label{tab:cqp}
\end{table*}

\section{Conversion from 4PCF to aperture statistics}
\subsection{Derivation of filters}
\label{sec:mmmm_derivation}
The aperture for the 4th order without conjugated shear is
\begin{eqnarray}
\begin{aligned}
    \mmmm &=\int  \diff^2 X_0 \cdots \int  \diff^2 X_3\, \left(\prod_{i=0}^3  Q_\theta(|\X{i}|)\frac{\breve{X}_i^{*^2}}{|\breve{X_i}|^2} \right)\, \Ga{0}^\m{cart}(\X{1}-\X{0},\X{2}-\X{0},\X{3}-\X{0})\\
    &= \int  \diff^2 \vt_1 \cdots \int  \diff^2 \vt_3 \int \diff^2 X_0\, \Ga{0}^\m{cart}(\bvt_1,\bvt_2,\bvt_3)\, \frac{1}{\left(4\pi\theta^4\right)^4}\, \left[\prod_{i=1}^3\left(\breve{X}_0^*+\vth{i}^*\right)^2\right]\, \breve{X}_0^{*2}  \exp{-\frac{|\X{0}|^2+\sum_{i=1}^3|\X{0}+\bvt_i|^2}{2\theta^2}},
\end{aligned}
\end{eqnarray}
where we chose the Cartesian projection for the shear and in the last step we changed to a reference system centred in $\X{0}$ and used the filter by \citet{crittenden2002} from \Eq{eq:filter_crit}. The polynomial in the exponent can be rewritten as $4(\X{0}-\q{0})^2+\sum_{i=0}^3(|\q{i}|^2)$, so if we change the integration over $\X{0}$ to the centre of mass of the quadrilateral, where $\X{0} = \bb{Z}+\q{0}$, in polar coordinates
\begin{align*}
    \mmmm =\frac{1}{\left(4\pi\theta^4\right)^4}\,  \int  \diff^2 \vt_1 \cdots \int  \diff^2 \vt_3\, \Ga{0}^\m{cart}(\bvt_1,\bvt_2,\bvt_3) \int \diff Z\, Z\, \exp{-\frac{4Z^2+\sum_{i=0}^3|\q{i}|^2}{2\theta^2}} \int \diff \varphi_Z\, \prod_{i=0}^3\left[\left(\breve{Z}^*+\qh{i}^*\right)^2\right].
\end{align*}

Both the integrals over $\varphi_Z$ and $Z$ can be carried out analytically, giving
\begin{eqnarray}
\begin{aligned}
    \mmmm &=\frac{1}{\left(4\pi\theta^4\right)^4}\,  \int  \diff^2 \vt_1 \cdots \int  \diff^2 \vt_3\, \Ga{0}^\m{cart}(\bvt_1,\bvt_2,\bvt_3)\,  \left(\prod_{i=0}^3\qh{i}^{*2}\,\exp{-\frac{|\q{i}|^2}{2\theta^2}}\right)\,\frac{2\pi \theta^2}{4} \\
    &= \int  \frac{\diff \vt_1 \ \vt_1}{\theta^2} \cdots \int  \frac{\diff \vt_3 \ \vt_3}{\theta^2}  \int  \frac{\diff\psi_{12}}{2\pi} \int  \frac{\diff\psi_{23}}{2\pi}\, K_0\left(\frac{\q{0}}{\theta},\frac{\q{1}}{\theta},\frac{\q{2}}{\theta},\frac{\q{3}}{\theta}\right)\, P_{0}^{\m{cart},\times}\left(\bvt_1, \bvt_2, \bvt_3\right)\,\Ga{0}^\m{\times}(\vt_1, \vt_2, \vt_3 , \psi_{12},\psi_{23}),
\end{aligned}    
\end{eqnarray}
where we defined the filter function $K_0$ from \Eq{eq:k0} by integrating over $\varphi_1$, since both the shear 4PCF and the combination of filter and projector are equivalent under a total rotation of a quadrilateral. Following the definition from \secref{sec:4PCF}, $P_{\mu}^{\m{cart},\times}\left(\bvt_1, \bvt_2, \bvt_3\right)= \left[P_{\mu}^{\times,\m{cart}}\left(\bvt_1, \bvt_2, \bvt_3\right)\right]^*$.

\medskip

When one of the vertices of the quadrilateral has a conjugated shear
    \begin{align}
        \mmmms &=\int  \diff^2 X_0 \cdots \int  \diff^2 X_3\,\prod_{i=0}^3 \left[ Q_\theta(|\X{i}|) \right]\,\prod_{j=1}^3 \left(\frac{\breve{X}_j^{*2}}{|\breve{X_j}|^2} \right)\,\frac{\breve{X_0}^2}{|\breve{X_0}|^2}\, \Ga{1}^\m{cart}(\X{1}-\X{0},\X{2}-\X{0},\X{3}-\X{0}) \nonumber \\
        &\hspace*{-1cm}=\frac{1}{\left(4\pi\theta^4\right)^4}\,  \int  \diff^2 \vt_1 \cdots \int  \diff^2 \vt_3 \,\Ga{1}^\m{cart}(\bvt_1,\bvt_2,\bvt_3)  \int \diff Z\,Z\,   \exp{-\frac{4Z^2+\sum_{i=0}^3|\q{i}|^2}{2\theta^2}} \int \diff \varphi_Z\, \prod_{j=1}^3\left[(\breve{Z}^*+\qh{j}^*)^2\right]\,(\breve{Z}+\qh{0})^2 \nonumber \\
        &\hspace*{-1cm}= \int  \frac{\diff \vt_1 \ \vt_1}{\theta^2} \cdots \int  \frac{\diff \vt_3 \ \vt_3}{\theta^2}  \int  \frac{\diff\psi_{12}}{2\pi} \int  \frac{\diff\psi_{23}}{2\pi}\, K_1\left(\frac{\q{0}}{\theta},\frac{\q{1}}{\theta},\frac{\q{2}}{\theta},\frac{\q{3}}{\theta}\right)\, P_{1}^{\m{cart},\times}\left(\bvt_1, \bvt_2, \bvt_3\right)\,\Ga{1}^\m{\times}(\vt_1, \vt_2, \vt_3 , \psi_{12},\psi_{23}),
    \end{align}
where we proceeded as for $\mmmm$.

When two of the vertices have conjugated shear
    \begin{align}
        \mmmsms &=\int  \diff^2 X_0 \cdots \int  \diff^2 X_3\, \prod_{i=0}^3 \left(  Q_\theta(|\X{i}|) \right)\,\frac{\breve{X}_1^{*2}}{|\breve{X_1}|^2}\frac{\breve{X}_2^{*2}}{|\breve{X_2}|^2} \frac{\breve{X_3}^2}{|\breve{X_3}|^2}\frac{\breve{X_4}^2}{|\breve{X_4}|^2}\, \Ga{5}^\m{cart}(\X{0}-\X{1},\X{2}-\X{0},\X{3}-\X{0}) \nonumber \\
        &\hspace*{-1cm}=\frac{1}{\left(4\pi\theta^4\right)^4}\,  \int  \diff^2 \vt_1 \cdots \int  \diff^2 \vt_3\, \Ga{5}^\m{cart}(\bvt_1,\bvt_2,\bvt_3)  \int \diff Z\, Z\,    \exp{-\frac{4Z^2+\sum_{i=0}^3|\q{i}|^2}{2\theta^2}} \nonumber \\
        &\hspace*{-1cm}\times \int \diff \varphi_Z\, (\breve{Z}^*+\qh{1}^*)^2(\breve{Z}^*+\qh{2}^*)^2(\breve{Z}+\qh{3})^2(\breve{Z}+\qh{4})^2 \nonumber \\
        &\hspace*{-1cm}=\int  \frac{\diff \vt_1 \ \vt_1}{\theta^2} \cdots \int  \frac{\diff \vt_3 \ \vt_3}{\theta^2}  \int  \frac{\diff\psi_{12}}{2\pi} \int  \frac{\diff\psi_{23}}{2\pi}\, K_5\left(\frac{\q{0}}{\theta},\frac{\q{1}}{\theta},\frac{\q{2}}{\theta},\frac{\q{3}}{\theta}\right)\,P_{5}^{\m{cart},\times}\left(\bvt_1, \bvt_2, \bvt_3\right)\, \Ga{5}^\m{\times}(\vt_1, \vt_2, \vt_3 , \psi_{12},\psi_{23}).
    \end{align}
A separation into $E$- and $B$-modes is possible following the prescription in \appref{app:NthOrderSeparation}, with the explicit separation in \secref{sec:map4_from_m4}.

\subsection{Visual inspection of filters for fourth-order aperture statistics}
\label{sec:filters_plot}
\begin{figure*}[htbp]
    \centering
    \includegraphics[height=.45\textwidth,valign=t]{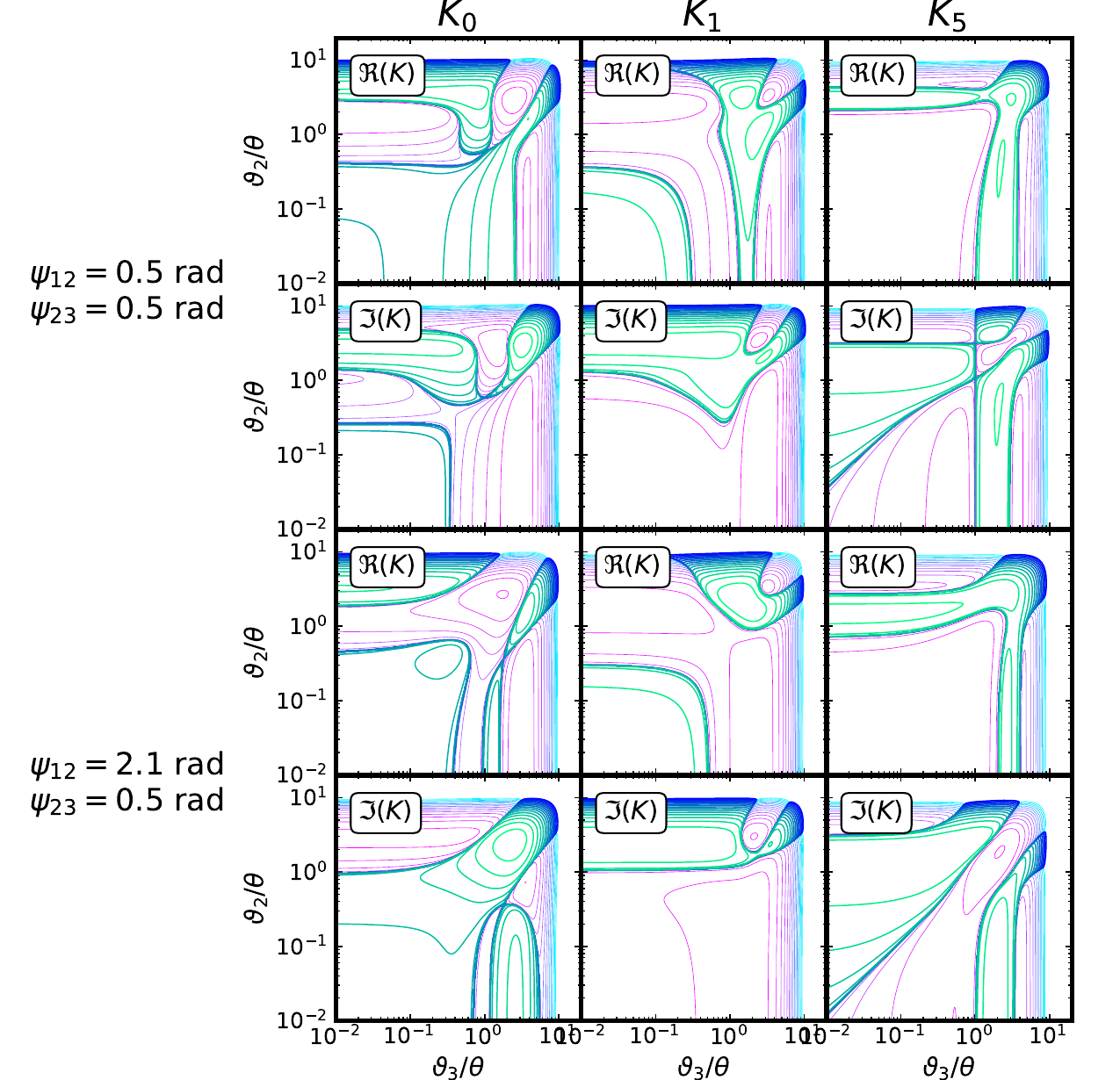}
    \centering
    \includegraphics[height = .45\textwidth,valign=t]{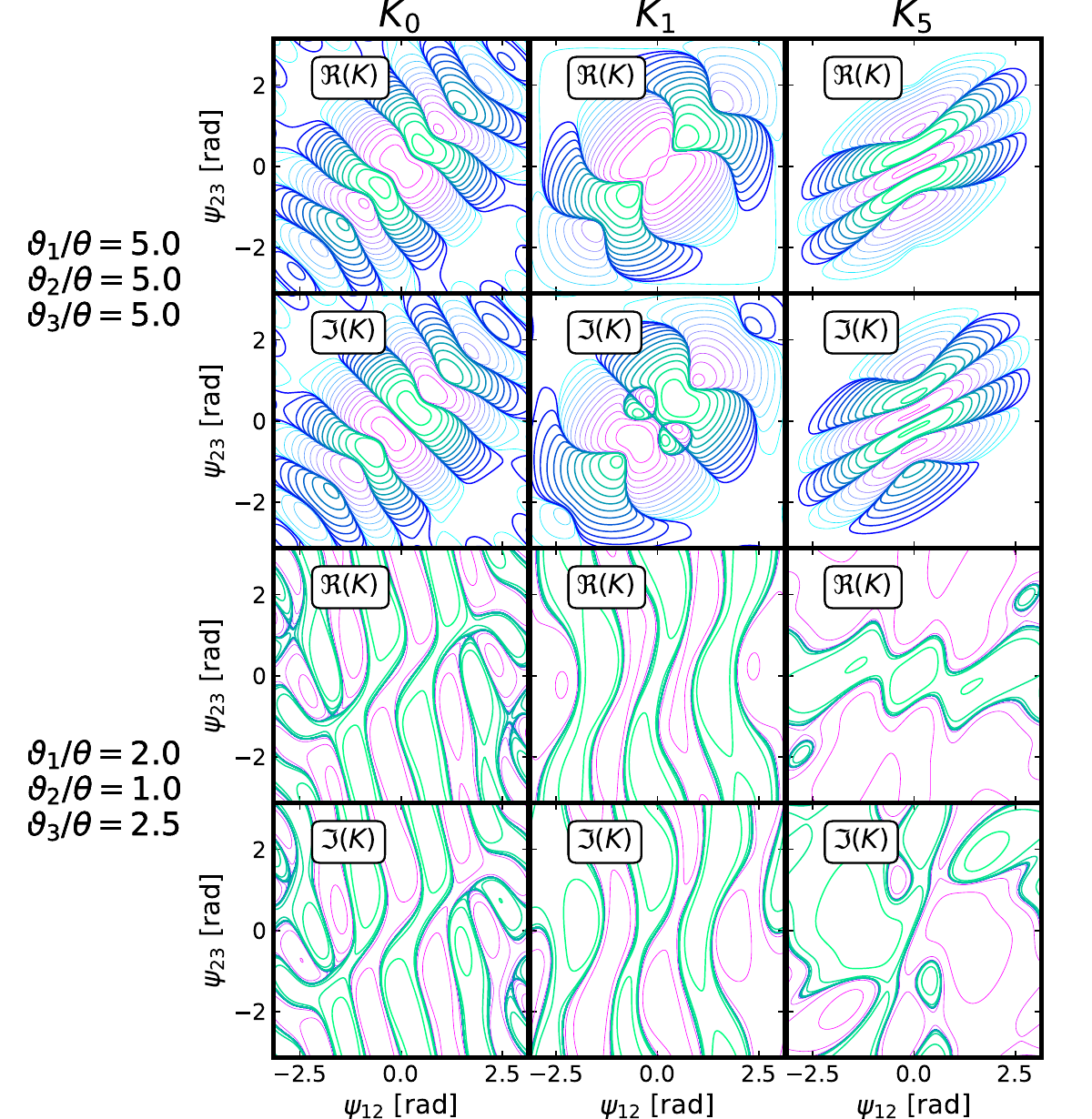}
  \caption{Filters for computing $\mmmm$, $\mmmms$, and $\mmmsms$ in terms of the 4PCF, as in         \Eq{eq:apertures_compute}; (Left) for given $\psi_{12}$ and $\psi_{23}$ combinations with the third radial variable fixed at $\vt_3 = \theta$  and (Right) for given radial combinations. Logarithmic contours with the minimum at $10^{-15}$ and a factor of 5 between them. Purple contours are positive values, green contours are negative ones.}
  \label{fig:filters}
\end{figure*}

In order to obtain insight into the shape and smoothness of the filters, the left panel in \Fig{fig:filters} shows the filter for selected angular combinations. In both the case of symmetric angles (first two rows) and in a more general case, the profiles present substructure at a wide range of scales and a Gaussian decay at radii $\vt\gg \theta$. 

The right panel in \Fig{fig:filters} shows the angular dependence of the filters for two radial combinations. For large radii (first two rows) the filters $K_i$ are peaked around $\psi_{12}=0=\psi_{23}$, while for $\vt_i \approx \theta$ the filters are smoother and alternate positive and negative. 

\section{Gaussian random fields: fourth-order statistics in terms of second-order statistics}
    \subsection{Correlation functions}
    \label{sec:4PCFas2PCF}

    For a GRF the 4PCF only contains disconnected terms, so it can be split into a combination of 2PCF \citep{isserlis1918}
    \begin{align}
\begin{split}
    \langle\g^{(*)}(\X{0})\g^{(*)}&(\X{1})\g^{(*)}(\X{2})\g^{(*)}(\X{3})\rangle = \langle\g^{(*)}(\X{0})\g^{(*)}(\X{1})\rangle\, \langle\g^{(*)}(\X{2})\g^{(*)}(\X{3})\rangle  \\
    &+ \langle\g^{(*)}(\X{0})\g^{(*)}(\X{2})\rangle\, \langle\g^{(*)}(\X{1})\g^{(*)}(\X{3})\rangle + \langle\g^{(*)}(\X{0})\g^{(*)}(\X{3})\rangle\, \langle\g^{(*)}(\X{1})\g^{(*)}(\X{2})\rangle,
\end{split}
    \label{eq:chi2}
\end{align}
    where the shear on each point can be either conjugated or not. Every 2PCF can be expressed as
    \begin{equation}
        \langle\g(\bb{X})\g(\bb{X} + \bb{s})\rangle = \frac{\breve{s}^4}{|\breve{s}|^4}\,\left[\xi_-(|\breve{s}|)+2\ii\xi_\times (|\breve{s}|)\right],\qquad \langle\g(\bb{X})\g^*(\bb{X} + \bb{s})\rangle = \xi_+(|\breve{s}|),\qquad \langle\g^*(\bb{X})\g^*(\bb{X} + \bb{s})\rangle = \frac{\breve{s}^{*4}}{|\breve{s}|^4}\,\left[\xi_-(|\breve{s}|)-2\ii\xi_\times (|\breve{s}|)\right]
    \end{equation}
    with $\breve{s}$ the complex number that represents the vector difference $\bb{s}$. For a field that does not violate parity $\xi_\times(|\breve{s}|) =0$ \citep{schneider2002b}. 
    Here we will explicitly keep $\xi_\times(|\breve{s}|)$ for the sake of completeness.

    In terms of the vectors $\bvt_i$ from \Fig{fig:cross_scheme} the fourth-order natural components are
    \begin{align}
        \begin{split}
            \Ga{0}^\m{cart}(\bvt_1,\bvt_2, \bvt_3) =&\left(\frac{\vth{1}}{|\vth{1}|}\right)^4 \left(\frac{\vth{3}-\vth{2}}{|\vth{3}-\vth{2}|}\right)^4\,\xi_-(|\vth{1}|)\, \xi_-(|\vth{3}-\vth{2}|) \\
            &+\ \left(\frac{\vth{2}}{|\vth{2}|}\right)^4\left(\frac{\vth{3}-\vth{1}}{|\vth{3}-\vth{1}|}\right)^4\,\xi_-(|\vth{2}|)\, \xi_-(|\vth{3}-\vth{1}|)\ +\ 
            \left(\frac{\vth{3}}{|\vth{3}|}\right)^4\left(\frac{\vth{2}-\vth{1}}{|\vth{2}-\vth{1}|}\right)^4\,\xi_-(|\vth{3}|)\, \xi_-(|\vth{2}-\vth{1}|),
            \label{eq:Gamma0cart} 
        \end{split}
    \end{align}
    \begin{align}
        \begin{split}
            \Ga{1}^\m{cart}(\bvt_1,\bvt_2, \bvt_3) =& \left(\frac{\vth{3}-\vth{2}}{|\vth{3}-\vth{2}|}\right)^4\,\xi_+(|\vth{1}|)\, \xi_-(|\vth{3}-\vth{2}|) \\
            &+\ \left(\frac{\vth{3}-\vth{1}}{|\vth{3}-\vth{1}|}\right)^4\,\xi_+(|\vth{2}|)\, \xi_-(|\vth{3}-\vth{1}|)\ +\ 
            \left(\frac{\vth{2}-\vth{1}}{|\vth{2}-\vth{1}|}\right)^4\,\xi_+(|\vth{3}|)\, \xi_-(|\vth{2}-\vth{1}|),
            \label{eq:Gamma1cart}
        \end{split}
    \end{align}
    \begin{align}
        \begin{split}
            \Ga{5}^\m{cart}(\bvt_1,\bvt_2, \bvt_3) =&\left(\frac{\vth{1}^*}{|\vth{1}|}\right)^4 \left(\frac{\vth{3}-\vth{2}}{|\vth{3}-\vth{2}|}\right)^4\,\xi_-(|\vth{1}|)\, \xi_-(|\vth{3}-\vth{2}|)\ +\ \xi_+(|\vth{2}|)\, \xi_+(|\vth{3}-\vth{1}|)\ +\ 
            \xi_+(|\vth{3}|)\, \xi_+(|\vth{2}-\vth{1}|) \ .        \label{eq:Gamma5cart}
        \end{split}
    \end{align}
    The natural components $\Ga{2-4}^\m{cart}$ can be obtained by renaming one of the vertices with a conjugated shear, for example $\Ga{2}^\m{cart}(\vth{1}, \vth{2}, \vth{3}) = \Ga{1}^\m{cart}(-\vth{1}, \vth{2}-\vth{1}, \vth{3}-\vth{1})$. $\Ga{6-7}^\m{cart}$, in turn, are equivalent to $\Ga{5}^\m{cart}$ when renaming the second vertex with conjugated shear, for example $\Ga{6}^\m{cart}(\vth{1}, \vth{2}, \vth{3}) = \Ga{5}^\m{cart}(\vth{2}, \vth{1}, \vth{3})$. This vertex exchange is an equivalent transformation to the one for the filters in \Eq{eq:trafo_filter}.

    These 4PCFs are finally converted to the $\times$ projection with the projector operators $P^{\times,\m{cart}}_l$.
    
    \subsection{Aperture measures}
    \label{sec:m4asm2}
    We start here from the aperture measures as defined in \Eq{eq:m4_base}, specifying to fourth-order and equal apertures.

    For $\mmmm$, when substituting \Eq{eq:chi2}, we find three integrals, which are identical under permutation of the integration variable names $\X{0}$ to $\X{3}$.
     Moreover, the eight-dimensional integrals can be factorized into two four-dimensional ones, which are also identical to each other, giving
    \begin{align*}
        \mmmm = 3\left[\int  \diff^2 X_0\,  Q_\theta(|\X{0}|)\int  \diff^2 X_1\,  Q_\theta(|\X{1}|)\frac{\breve{X}_0^{*2}}{|\breve{X}_0|^2}\frac{\breve{X}_1^{*2}}{|\breve{X}_1|^2}\langle\g(\X{0})\g(\X{1})\rangle\right]^2 = 3\left[\int  \frac{\diff |\breve{s}| |\breve{s}|}{\theta^2} \left(\xi_-(|\breve{s}|)+2\ii\xi_\times (|\breve{s}|)\right)T_-\left(\frac{|\breve{s}|}{\theta}\right)\right]^2,
    \end{align*}
    where in the last step we changed $\X{1} = \X{0} + \bb{s}$ and defined the functions $T_-$ as in \citet{jarvis2004}, where in the derivation the cross term was implicitly set to zero. Therefore $\mmmm = 3[\langle M^2\rangle(\theta)]^2$.

    For $\mmmms$, when substituting \Eq{eq:chi2}, the three integrals are identical under permutation of the integration variable names $\X{1}$ to $\X{3}$ and the eight-dimensional integrals can be factorized into two four-dimensional ones, i.e.
    \begin{align*}
        \mmmms &= 3\left[\int  \diff^2 X_0\,  Q_\theta(|\X{0}|)\int  \diff^2 X_1\, Q_\theta(|\X{1}|)\frac{\breve{X_0}^{2}}{|\breve{X_0}|^2}\frac{\breve{X_1}^{*2}}{|\breve{X_1}|^2}\langle\g^*(\X{0})\g(\X{1})\rangle\right]\\
        &\times \ \left[\int  \diff^2 X_2 \, Q_\theta(|\X{2}|)\int  \diff^2 X_3 \,Q_\theta(|\X{3}|)\frac{\breve{X_2}^{*2}}{|\breve{X_2}|^2}\frac{\breve{X_3}^{*2}}{|\breve{X_3}|^2}\langle\g(\X{2})\g(\X{3})\rangle \right]\\
        & = 3\left[\int  \frac{\diff |\breve{r}| |\breve{r}|}{\theta^2} \xi_+(|\breve{r}|)T_+\left(\frac{|\breve{r}|}{\theta}\right)\right]\left[\int  \frac{\diff |\breve{s}| |\breve{s}|}{\theta^2} \left[\xi_-(|\breve{s}|)+2\ii\xi_\times (|\breve{s}|)\right] T_-\left(\frac{|\breve{s}|}{\theta}\right)\right],
    \end{align*}
    where we rewrote the variables $\X{1} = \X{0} + \bb{r}$ and $\X{3} = \X{2} + \bb{s}$, and defined $T_+$ as in \citet{jarvis2004}. Together, $\mmmms = 3\left< M^2\right>(\theta)\left< MM^*\right>(\theta)$, and equivalent for the other aperture measures with one conjugated component.
    %Since $\left< MM^*\right>(\theta)$ is purely real, $\left< MM^*\right>(\theta) = \left< M^*M\right>(\theta)$ and $\left< M^*M^3\right>(\theta) =\left< MM^*M^2\right>(\theta)= \left< M^2M^*M\right>(\theta)= \left< M^3M^*\right>(\theta)$.

    Finally, for $\mmmsms$ only two of the three integrals obtained when substituting \Eq{eq:chi2} are identical under permutation of $\X{0}$ and $\X{1}$, or $\X{2}$ and $\X{3}$, giving
    \begin{align*}
    \mmmsms =&2\left[\int  \diff^2 X_0\,  Q_\theta(|\X{0}|)\int  \diff^2 X_2 \,Q_\theta(|\X{2}|)\frac{\breve{X_0}^{2}}{|\breve{X_0}|^2}\frac{\breve{X_2}^{*2}}{|\breve{X_2}|^2}\langle\g^*(\X{0})\g(\X{2})\rangle \right]^2 \\
    &+ \left\{ \left[\int  \diff^2 X_0 \, Q_\theta(|\X{0}|)\int  \diff^2 X_1 \,Q_\theta(|\X{1}|)\frac{\breve{X_0}^2}{|\breve{X_0}|^2}\frac{\breve{X_1}^2}{|\breve{X_1}|^2}\langle\g^*(\X{0})\g^*(\X{1})\rangle \right] \right. \\
    &\left. \times\ \left[\int  \diff^2 X_2 \, Q_\theta(|\X{2}|)\int  \diff^2 X_3 \,Q_\theta(|\X{3}|)\frac{\breve{X_2}^{*2}}{|\breve{X_2}|^2}\frac{\breve{X_3}^{*2}}{|\breve{X_3}|^2}\langle\g(\X{2})\g(\X{3})\rangle\right]\right\}\\
    =&2\left[\int  \frac{\diff |\breve{r}| |\breve{r}|}{\theta^2} \xi_+(|\breve{r}|)T_+\left(\frac{|\breve{r}|}{\theta}\right)\right]^2 +\mathrm{Abs}\left(\int  \frac{\diff |\breve{s}| |\breve{s}|}{\theta^2} \left(\xi_-(|\breve{s}|)+2i\xi_\times (|\breve{s}|)\right)T_-\left[\frac{|\breve{s}|}{\theta}\right]\right)^2,
    \end{align*}
    where we proceeded as in the previous cases and used that complex conjugation is distributive over addition and multiplication. Then $\mmmsms = 2\left[\left< MM^*\right>(\theta)\right]^2 + \mathrm{Abs}\left[\left< M^2\rangle(\theta)\right>\right]^2$ and $\left< M^*MM^*M\right>(\theta) = \left< M^*M^{2}M^*\right>(\theta) =\left< M^{*2}M^{2}\right>(\theta)$.

\twocolumn
\section{Estimation of derivatives from samples}
\label{app:derivatives}
\subsection{Method}
\label{app:method}
If we have access to a cosmological simulation suite with $i=\{1, \cdots, N_\m{cosmo}\}$ distinct sets of cosmological parameters or samples, $\bb{\mu}^i$, we can measure an observable, $y = f(\bb{\mu})$, on the different cosmologies, $y^i = f(\bb{\mu}^i)$.\footnote{We use upper indices to label each cosmological set and lower indices for the different cosmological parameters on each set.} Here we approximate $f$ by a first-order Taylor expansion around the fiducial values of the $N_\m{param}$ parameters,
\begin{align}
    f(\bb{\mu}) &= \left.f\right|_\m{fid} + \sum_{j=1}^{N_\m{param}} \left.\frac{\partial f}{\partial \mu_j}\right|_\m{fid}\left(\mu_j-\mu_{j,\m{fid}}\right) + \mathcal{O}(\bb{\mu}-\bb{\mu}_\m{fid})^2 \\ &\approx b + \sum_{j=1}^{N_\m{param}} a_{j}\left(\mu_j-\mu_{j,\m{fid}}\right);\label{eq:model_parameters} \\
    f(\bb{\mu}^i)  &= b + \sum_{j=1}^{N_\m{param}} a_{j}\left(\mu_j^i-\mu_{j,\m{fid}}^i\right) \ ,
    \label{eq:taylor}
\end{align}
where in the first step we introduced the model parameters, $a_{j}$ and $b$, and in the second step we specified for every sample cosmology. By defining the quantity $\bb{\mu}^{'i} = \left(\mu_{1}^i-\mu_{1,\mathrm{fid}}^i, \cdots , \mu_{N_\m{param}}^i-\mu_{N_\m{param},\mathrm{fid}}^i, 1\right)$ we can rewrite \Eq{eq:taylor} as the linear combination
\begin{equation}
    f(\bb{\mu}^i) = \sum_{j=1}^{N_\m{param}+1} a_{j}^{'}\mu_j^{'i},
\end{equation}
with model parameters $a_{j}^{'} = a_{j}$ if $j \leq N_\m{param}$ and $a_{N_\m{param} +1}' = b$. The best-fit model parameters are such that they minimize the loss function,
\begin{align}
    \label{eq:lossa}
    &\mathcal{L} = \sum_{i=1}^{N_\m{cosmo}} w^i |y^i - f(\bb{\mu}^i)|^2, \quad \m{with} \\
    &w^i = \exp{-\frac{1}{2\sigma^2}\left(\bb{\mu}^i-\bb{\mu}|_\m{fid}\right)^{\rm T} \tens{C}_{\mu,\mu}^{-1}\left(\bb{\mu}^i-\bb{\mu}|_\m{fid}\right)},
    \label{eq:loss}
\end{align}
where $w^i$ is the weight of the $i$-th set of cosmological parameters. The $w^i$ depend on the distance in cosmological parameter space to the the fiducial cosmology; on the covariance matrix of the cosmological parameters, $\tens{C}_{\mu,\mu}$, which is expected to be diagonal if the cosmologies are distributed on a latin hypercube, and can be computed from the set of cosmological parameters; and on the hyperparameter $\sigma^2$, the weighting scale. With $w^i$ we give a higher weight to sets of cosmological parameters that are closer to the fiducial value, where the linear approximation for $f$ is more accurate, but also more prone to noise.

The minimization of \Eq{eq:lossa} w.r.t to the parameter $a'_k$
\begin{equation}
    \frac{\partial \mathcal{L}}{\partial a'_k}=0 \rightarrow \sum_{i=1}^{N_\m{cosmo}} w^i y^i \mu^{'i}_k = \sum_{j=1}^{N_\m{param}+1}\sum_{i=1}^{N_\m{cosmo}} w^i \mu^{'i}_k \mu^{'i}_j a'_j,
\end{equation}
or in matrix formalism for all model parameters, defining $(M_w)^i_j= w^i \mu^{'i}_j$ and $(M)^i_j= \mu^{'i}_j$,
\begin{equation}
    \bb{M_w}^T \bb{y} = \bb{M_w}^T \bb{M} \bb{a}' \rightarrow \bb{a}' = (\bb{M_w}^T \bb{M})^{-1}\bb{M_w}^T \bb{y}.
\end{equation}
From \Eq{eq:model_parameters} we see that the partial derivatives are approximated by the first $N_\m{param}$ components of the model parameters, $\bb{a}'$, while $a_{N_\m{param}+1}'$ is an approximation of the observable at the fiducial cosmology. 

Due to the construction of the weights, considering in the analysis a cosmological parameter on which the observables depend weakly, introduces noise in the derivatives w.r.t. all parameters. In such case, the derivatives w.r.t. the parameters with strong dependency can be made more stable by marginalizing over the noisy parameters.

\subsection{Derivatives from the \textit{cosmo}-SLICS ensemble}
 \label{app:derivatives_cS}
 \begin{figure*}[htbp]
    \centering
    \includegraphics[width=.9\textwidth,valign=t]{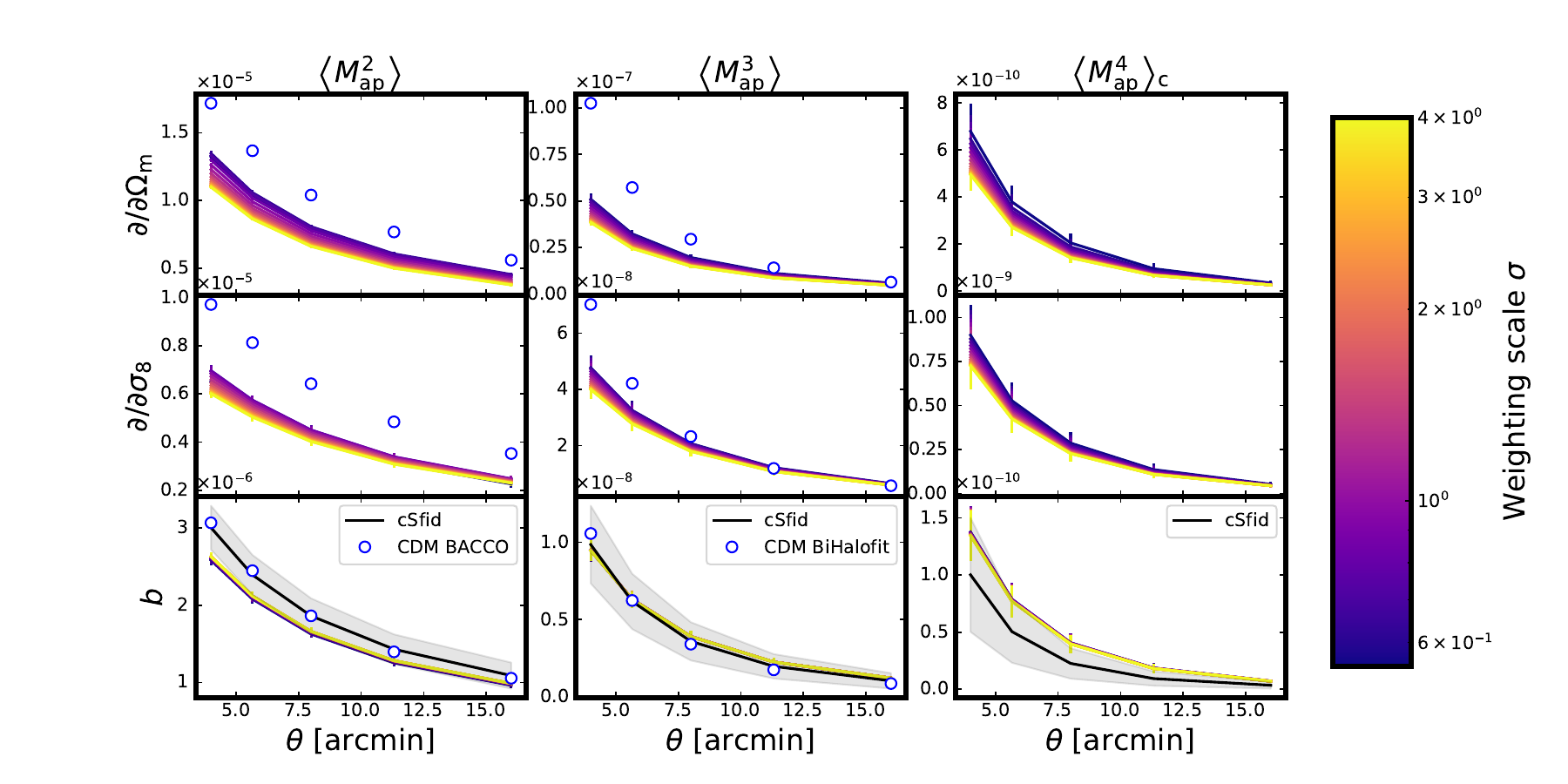}
  \cprotect\caption{Stability of the model parameters for the derivatives when varying the weighting scale $\sigma$ (see \Eq{eq:loss}). All points have error bars from Jackknifing over light cones. The first two rows correspond to the derivatives with respect to the cosmological parameters and the third one to the model prediction for the given order of the aperture statistics. Blue rings mark the values obtained from theory for the aperture statistics and their derivatives, estimated from finite differences. The agreement between $b$ and the measurement in \verb|cSfid| (black) is a test of the robustness of the method.}
  \label{fig:derivatives}
\end{figure*}
Since the derivatives only depend on one hyperparameter, $\sigma$, we can study the stability of the derivatives by inspecting their dependency on $\sigma$, considering \verb|cSfid| as fiducial cosmology. From the cosmological parameters considered in \textit{cosmo}-SLICS (cS19),\cprotect\footnote{From the 26 sets of cosmological parameters available, each one of them with 25 lightcones, we discard the third light cone, as one of the redshift slices has a significant $\langle M_\m{ap} \rangle \neq 0$ and discarding only such map would change the $n(z)$.} %Moreover, the proximity in parameter space of T17 and the cosmology \verb|fid| in cS19, cS\verb|fid|, results in a weight too large for cS\verb|fid|, increasing the global noise in the derivatives, so we discard it in our analysis.}
 $\mu = \{\Omegam, S_8, h, w_0\}$, we marginalize over $h$ and $w_0$, since we find that the non-tomographic aperture statistics on cS19 do not have enough constraining power to produce robust derivatives on such parameters. 

The first two panels in \Fig{fig:derivatives} show the derivatives of $\left<M_\m{ap}^2\right>$, $\left<M_\m{ap}^3\right>$, and $\left<M_\m{ap}^4\right>_c$, with Jackknifing error bars over light cones and colour coded by their weighting scale $\sigma$. For small $\sigma$ only a few points contribute significantly, so the derivatives are biased towards the noisy estimate from a few cosmologies close to the fiducial value. The derivatives converge for increasing $\sigma$ to the value we would recover if we set all weights to unity, therefore far from the linear approximation regime. 

This study on $\sigma$ is equivalent to investigating the case where the derivatives are only computed with the $A\leq N_\m{cosmo}$ cosmologies closest to the fiducial cosmology, where we obtain convergence for $A>18$ with $\sigma=0.6$, and noise-dominated derivatives otherwise. The continuous nature of varying $\sigma$ makes the noise for $\sigma < 0.6$ appear as a smooth bias in the derivatives.

For second- and third-order statistics we can compute the derivatives with finite differences from a theoretical model, with the values marked as blue rings in \Fig{fig:derivatives}.\cprotect\footnote{We compute second-order statistics from the non-linear cold-dark-matter BACCO emulator \citep{nonlinearbacco2021} and \verb|pyccl| \citep{pyccl} and third-order statistic from the cold-dark-matter BiHalofit \citep{bihalofit}.} Our method underestimates the derivatives by a degree that decreases for decreasing $\sigma$ until the noise limit of $\sigma\approx 0.6$ in cS19. 

We generated theoretical second- and third-order statistics that correspond to cosmological parameters close to \verb|cSfid| in the $\{\sigma_8,\Omegam\}$ plane and used them to compute derivatives with the loss minimization method, recovering the theoretical derivatives from finite differences. The bias between the finite differences method and the loss minimization method in \Fig{fig:derivatives} therefore arises from the sparsity of cS19 around \verb|cSfid| and the linearity assumption from \Eq{eq:taylor}. With these theoretical (noiseless) points close to \verb|cSfid| we find the best results for a combination of $N_\m{cosmo}<25$ and $\sigma \ll 1$. When studying only cS19, such combinations are dominated by noise, so we proceed with $N_\m{cosmo} = 25$, $\sigma = 0.6$.
%Due to the sparsity of cS19 around the fiducial cosmology in the $\{\sigma_8,\Omegam\}$ plane, weighting scales $\sigma<0.6$ are noise-dominated and therefore not available. 
%Upon visual inspection of the agreement between theory and our model we set for the forecast $\sigma=2$, which lies within the stable regime for second-order statistics.

The third panel in \Fig{fig:derivatives} shows the predicted value of the observable, $b$, along with the value measured on \verb|cSfid| (in black) and the theoretical model for second- and third-order statistics (blue rings). We find that, with our method, we predict a value of the observables that is compatible within $1\sigma$ to the one measured in \verb|cSfid|. We attribute the difference in second- and fourth-order statistics to the slight changes in redshift distribution for the different cS19 cosmological sets, arising from the varying number of convergence maps for each set. Moreover, neighboring aperture scales are highly correlated, which is not considered in the Jackknifing error bars nor in the $1\sigma$ region from \verb|cSfid|.

In summary, we find that derivatives are stable when considering the cosmological parameters $\Omegam$ and $\sigma_8$ and marginalizing over $h$ and $w_0$, but become unstable if we try to constrain all four parameters simultaneously. When using sparse simulations around the fiducial value, as cS19 around \verb|cSfid|, the derivatives are biased towards lower values, resulting in overestimated Fisher ellipses for the independent aperture statistics, and possibly under- or overestimated for the joint analysis. Even on such sparse simulations, we recover a value of the mean for second- and third-order statistics compatible within their $1\sigma$ precision to the measured one.

%We find that, with our method, we predict a value of the observables that is compatible within $1\sigma$ to the one measured on T17, albeit the different assumptions that go into the cS19 and T17 simulations. The robustness of our method is further tested by predicting the cosmology cS\verb|fid| from the other cosmologies in cS19, with similar agreement.

%The third panel in \Fig{fig:derivatives} shows the predicted value of the observable, $b_\m{T17}$, along with the value measured on T17. The fourth panel, in turn, shows the prediction, $b_\m{fid}$ when we use the cosmology labeled as \verb|fid| from cS19 (cS\verb|fid|) along with the measured value on cS\verb|fid|. T17 and cS\verb|fid| have very similar cosmological parameters, from which we deduce that the discrepancy on $b_\m{T17}$ arises from the different assumptions between T17 and cS19, rather than a methodological issue, i.e. $b_\m{T17}$ is the prediction from cS19 at the cosmology of T17.

%With the method described in \secref{app:derivatives} the values of the observables at the fiducial cosmology are not directly used in the computation of the derivatives, and $b$ is only used as a test, so we proceed with the fiducial cosmology from T17. 

\subsection{Stability of the ellipses}
 \label{app:stability_ellipses}
 
Fisher ellipses are fully described by two quantities, i.e. a direction and an area. For the two-dimensional distribution of the $i$-th and the $j$-th parameters, these quantities can be the Pearson correlation coefficient,
\begin{equation}
    p_{i,j} = \frac{C_{i,j}}{\sqrt{C_{i,i}C_{j,j}}},
\end{equation}
with $C = F^{-1}$ and $F$ the Fisher matrix, and the Figure of Merit of order $n$,
\begin{equation}
    \m{FoM}_{i,j} = \left[2\pi n^2\ \left(C_{i,i}+C_{j,j}\right)\ \sqrt{1-p_{i,j}^2}\right]^{-1},
    \label{eq:FoM}
\end{equation}
which is the inverse of the area of the ellipse enclosing the $n\sigma$ region, where here $n=1$.

\begin{figure*}[htbp]
    \centering
    \includegraphics[width=1\textwidth,valign=t]{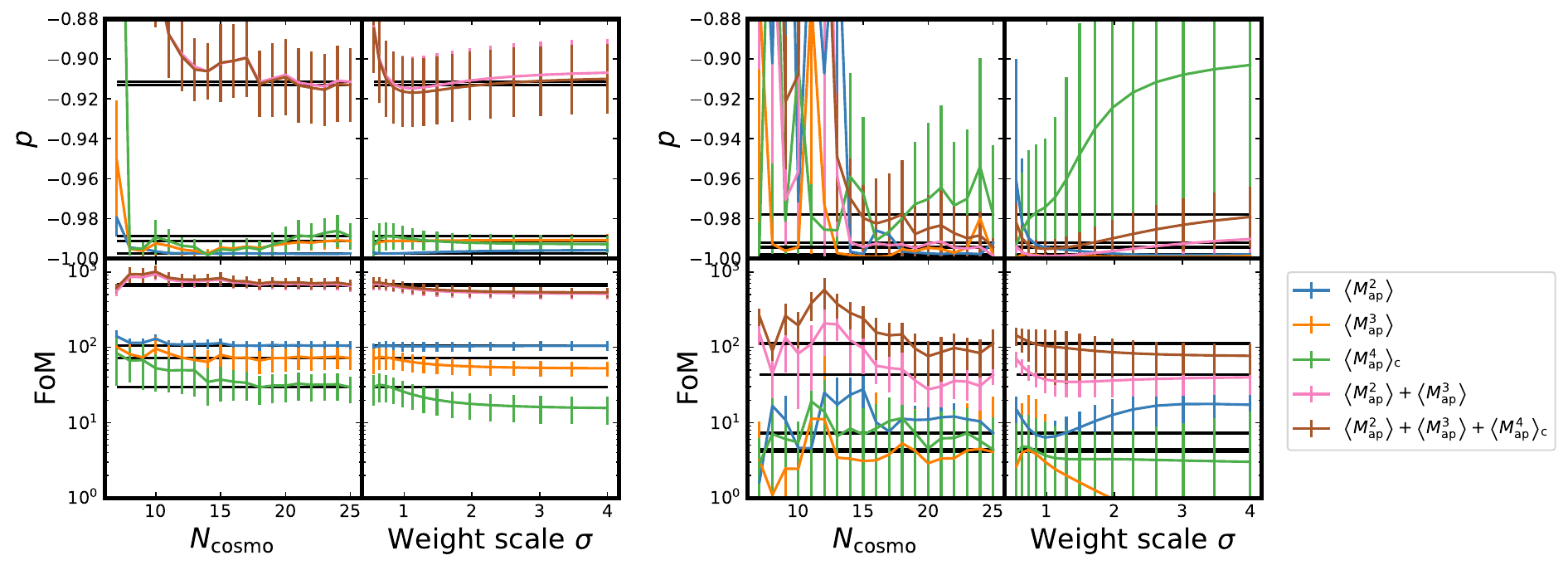}
    \centering
  \caption{Stability of the $\{\Omegam,\sigma_8\}$ Fisher ellipses when varying the number of cosmologies around the fiducial value, $N_\m{cosmo}$, or the weighting scale, $\sigma$; specified through the Pearson correlation coefficient $p$, measure of the orientation of the ellipse, and the Figure of Merit FoM, measure of the area of the ellipse. The different colours correspond to the given combination of order for the aperture statistics considered in the analysis. Comparison of the results when marginalizing over $h$ and $w_0$ in the computation of the ellipses (Left), with stable ellipses; and the results when we try to constrain all cosmological parameters from cS19 (Right), where projection effects from the unconstrained parameters make the ellipses unstable.}
  \label{fig:ellipses}
\end{figure*}

The left panel in \Fig{fig:ellipses} shows the stability of $p$ and the $\m{FoM}$ when varying the number of cosmologies considered for the derivatives or the weighting scale. When using too few cosmologies, or $\sigma <0.6$, we do not capture enough information, and the ellipses are noise-dominated. The horizontal lines, using all $N_\m{cosmo}=25$ and $\sigma=0.6$, are the values used for the Fisher forecast. $p$ and the $\m{FoM}$ are compatible within their Jackknifing error bars for $\{\left<M_\m{ap}^2\right>,\left<M_\m{ap}^3\right>\}$ and $\{\left<M_\m{ap}^2\right>,\left<M_\m{ap}^3\right>,\left<M_\m{ap}^4\right>_c\}$, so we do not report a significant improvement when including fourth-order statistics, as discussed in \secref{sec:fisher}.

The right panel in \Fig{fig:ellipses} shows the equivalent analysis on the $\{\Omegam,\sigma_8\}$ ellipse, when we do not marginalize over $h$ and $w_0$ in the computation of the derivatives. The $\m{FoM}$ is around an order of magnitude smaller, sign that we cannot constrain all 4 cosmological parameters. Albeit noise-dominated, we find the strongest change for fourth-order statistics, hinting towards a stronger dependency of fourth-order statistics on $h$ and $w_0$.

\end{appendix}
\end{document}